# How sure are we? Two approaches to statistical inference

Michael Wood
University of Portsmouth, UK
michaelwoodslg@gmail.com or michael.wood@port.ac.uk
http://woodm.myweb.port.ac.uk/SL/statistics.htm

*10 March, 2018*

## Abstract

Suppose you are told that taking a statin will reduce your risk of a heart attack or stroke by 3% in the next ten years, or that women have better emotional intelligence than men. You may wonder how accurate the3% is, or how confident we should be about the assertion about women's emotional intelligence, bearing in mind that these conclusions are only based on samples of data? My aim here is to present two statistical approaches to questions like these. Approach 1 is often called null hypothesis testing  but I prefer the phrase "baseline hypothesis": this is the standard approach in many areas of inquiry but is fraught with problems. Approach 2 can be viewed as a generalisation of the idea of confidence intervals, or as the application of Bayes' theorem. Unlike Approach 1, Approach 2 provides a tentative estimate of the probability of hypotheses of interest. For both approaches, I explain, from first principles, building only on "common sense" statistical concepts like averages and randomness, both how to derive answers, and the rationale behind the answers. This is achieved by using computer simulation methods (resampling and bootstrapping using a spreadsheet available on the web) which avoid the use of probability distributions (t, normal, etc). Such a minimalist, but reasonably rigorous, analysis is particularly useful in a discipline like statistics which is widely used by people who are not specialists. My intended audience includes both statisticians, and users of statistical methods who are not statistical experts.

*This paper makes use of two spreadsheets:*
     http://woodm.myweb.port.ac.uk/SL/resample.xlsx  for resampling and bootstrapping and calculating probabilities. (There are several versions of this with different data sets and menu options chosen.)
     http://woodm.myweb.port.ac.uk/CLIP.xls  for estimating confidence or probability levels from confidence intervals and p values.

# Contents

# Introduction

If you want to find something out about the world, the first step is to get some data - look at what is happening, do an experiment or whatever. Then try to come to a conclusion. The difficulty with which statistics grapples is that your conclusions may be wrong or inaccurate because of a multitude of unknown factors, or because of "chance". I went along to my doctor recently with some good news about a sore ankle. The painkillers I had been using seemed to go on working even when I stopped taking them. Brilliant! But she just gave me this pitying, patronizing look, and explained that sore ankles come and go for reasons that nobody understands. The painkillers had nothing to do with my miraculous cure: it was a matter of pure "chance". Or, to put it another way, she didn't know what were the hidden factors that had made my ankle better, but she was pretty sure it wasn't the painkillers.

How could she be so sure? (The subsequent history of the ankle suggests she was right.) That's the problem I want to tackle here. It's a common problem, to which a lot, perhaps most, of statistical theory is devoted to tackling.

First, a brief aside about my approach here. The difficulty with a lot of statistical theory is that it's difficult: the concepts are subtle, the methods complicated, the maths horrendous, and the conclusions are often need an expert to interpret. My aim here is to present a rather different version which is as short and user-friendly as possible, but still as powerful and useful as the standard version, perhaps even more so. (I have explained how my approach relates to the standard concepts here.) Above all, I want it to make sense: no mysterious formulae, nothing which has to be accepted on trust.

Now back to the main story. There are two basic approaches to deciding how sure we are in a statistical sense. The first is to take a hypothesis as a baseline, and see how likely our data is assuming this hypothesis true. The second is to try to assess, using our data, the probability[1] of, or confidence level for, whatever is of interest to us. The first is relatively straightforward, but *does not* tell us directly how certain our conclusions are. The second approach is harder , but *does* produce clear, although usually tentative, answers to the question "How sure are we?"

## Approach 1: Is the data consistent with a baseline hypothesis?

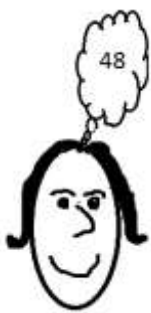

A few years ago I did an experiment to see if I could detect telepathy (communication between people without using any of the known senses - sight, hearing, touch, etc). I was in one room with a pack of 50 cards, numbered from 1 to 50, which I shuffled thoroughly, and then cut to display one of the cards which I showed to a volunteer - Peter - who concentrated hard on it. Then, in another room, I had another volunteer - Annette - who at the pre-arranged time, tried to see if she knew which of the 50 cards Peter was thinking about. She then wrote the number down on a piece of paper, and got it right - Number 48! She said afterwards that it felt as though she was guessing, but it was the number 48 which floated into her mind. We checked carefully, but there was definitely no way she could have known which card was chosen except by means of telepathy. (This, in common with most of the other examples below, isn't true[2]: fictional examples have the advantage that they can be designed to make the point as clearly as possible.)

There were only two viable explanations: either Annette was guessing and was lucky, or she was communicating telepathically with Peter. Which explanation is right?

The baseline hypothesis here is that Annette was guessing and her success was just due to chance. With 50 cards the probability of getting the answer right by chance is only 1/50 (2%). So chance is unlikely and telepathy is the more likely hypothesis. Or is it? You will probably have a few doubts about this conclusion!

Now suppose that there were a million cards and Annette still got the right one. The probability of getting it right by chance is now one in a million (0.0001%). If there really were no other explanations except chance and telepathy, which would you go for? Many people would say that with 50 cards chance is the more likely explanation, whereas with a million cards they would go for telepathy[3]. *The lower this probability is the greater the evidence against the chance hypothesis and for the telepathy hypothesis.*

## The shuffle test

Let's look at another example which is more like the sort of problem statistics regularly deals with. Is being a vegetarian good for your wellbeing? Are people who eat lots of vegetables and no meat happier and healthier than meat-eaters? To find out, a random sample of 60 people were given a wellbeing test which gave a score on a 0 (awful) to 100 (perfect) scale, and asked if they were a vegetarian (veggie) or an omnivore (omnie). I'll leave it to your imagination whether wellbeing encompasses physical health, mental health, happiness or whatever. A summary of the results is in Figure 1, and all the data and analysis is at
http://woodm.myweb.port.ac.uk/SL/resample60baselinetest.xlsx. (There are different ways in which this comparison might be organised: you might take separate samples of veggies and omnies, or you might take a mixed sample and tell some of them to adopt a veggie diet and the rest an omnie diet - this is the only way of seeing if vegetarianism *causes* enhanced wellbeing . I discuss this [below](#).)

*Figure 1. Some of the well-being data from 60 people*

| Wellbeing score | Diet |
|---:|:---|
| 74 | Vegetarian |
| 65 | Vegetarian |
| 69 | Omnivore |
| 37 | Omnivore |
| 57 | Vegetarian |
| 26 | Omnivore |
| ... 54 rows of data omitted ... | |
| Average score for 27 veggies | = 60 |
| Average score for 33 omnies | = 50 |

There were 27 vegetarians in the sample with an average[4] score of 60, and 33 omnivores with an average score of 50. The veggies' average score is 10 units higher than the omnies'. But the question now is whether this result is an accurate reflection of the difference between the *whole population* of vegetarians and the whole population of omnivores? Perhaps another sample would give a different result? The six rows of data in Fig 1, for example, give a more extreme result: the vegetarian average is 65 and the omnivore average is 44, a difference of 21.

To tackle this question let's see *what would happen if there really were no difference between the two groups*. If this were so, each of the six well-being scores in Table 1 would be equally likely to be from a vegetarian or an omnivore. All we do now is to run some experiments to see what is likely to happen if there really were no underlying difference between the two groups. The method is known as a randomization test, although I prefer the term shuffle test for reasons that should be obvious.

To do this for the six rows of data in Figure 1, I wrote each of the six well-being scores in Fig 1 on a card, shuffled the cards, and then dealt three in a vegetarian pile, and three in an omnivore pile. The results were:

*Figure 2. Six card shuffle to see what might happen if no difference between groups*

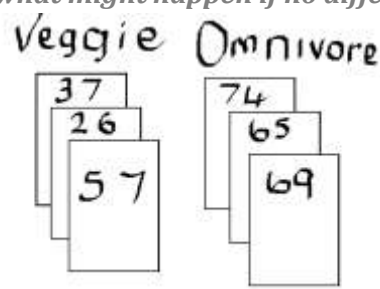

The average veggie score here is 40 and the omnie average is 69, a difference of 29 in the opposite direction from the real sample. The next shuffle produced a difference of 8 in favour of the vegetarians, and the next 24 in favour of the omnivores. Shuffling the data like this can produce wild variations in the results. *This suggests that the first three rows of data in Figure 1, with a difference between the groups of 21, could easily have arisen if there were no consistent difference between vegetarians and omnivores, and that the apparent difference between veggies and omnies is down to chance*.

However, the whole sample comprised 27 veggies and 33 omnies. This is a much larger sample which makes shuffling cards impractical, so it's helpful to use a computer. The spreadsheet http://woodm.myweb.port.ac.uk/SL/resample60baselinetest.xlsx is set up to do these shuffles and analyse the results. If you click on the Single resample tab at the bottom you will see one shuffle. All the computer is doing is shuffling the 60 wellbeing scores and dealing them at random to the 27 veggies and 33 omnivores.

These shuffles are called resamples because you are (sort of) taking a sample from the original sample. Press F9 (or Formulas-Calculate Now) and you will see the numbers rearranging corresponding to another shuffle. When I did this the first few results were -7, -5, 0, -3, ... (rounding off to the nearest whole number). Each of these is the difference between the veggie and omnie averages (veggie - omnie) for one resample. In the six card shuffle in Fig 2 the difference is 29 in favour of the omnivores, which the spreadsheet would call -29. Not surprisingly, the results from the very small sample of three in each group (-29, 8, -24) fluctuate much more than the results from the whole sample of 60.

Now click on the Lots of resample tab and you will see the results from 1000 shuffles. This should be similar to Figure 3 (but not identical because all shuffles are random).

*Figure 3. Histogram showing 1000 resamples (shuffles)*

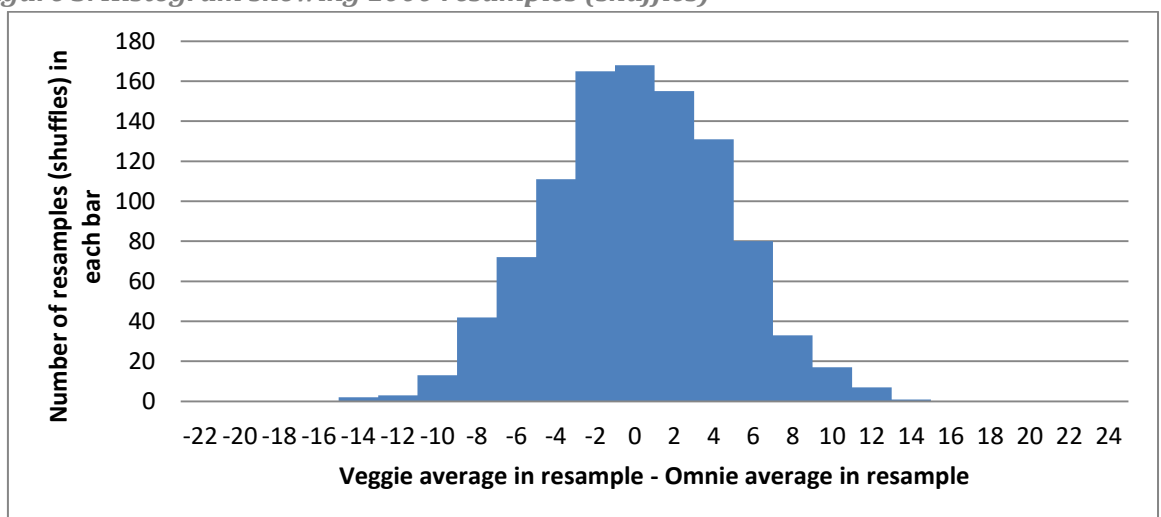

The bar centred on 0, for example, indicates that 166 of the 1000 resample averages lay between -1 and +1.

Figure 3 shows that the majority of the resamples produced a difference less than the difference in the actual sample (10), but a few were as high as +14 or as low as -14.

We can now count up the proportion of resamples which had a difference of 10 or more, or -10 or less. This will tell us how likely it is that results as extreme as the actual result (+10) could be produced by chance alone. When I did this the answer was 2%. In other words just 20 of 1000 resamples had a difference between the two groups of 10 or more or -10 or less. (You can check this by scrolling down the results in the pink cells in the Lots of resamples sheet.)

This suggests that if there were no consistent differences between veggies and omnivores, and any differences are down to chance, the probability of getting results with as big difference as in the sample of 60 is only about 2%. Which probably means that you would be tempted to think the evidence is quite strong that vegetarianism really does improve wellbeing.

This is a very similar to the telepathy experiment. The probability of getting the result - Annette choosing the correct card - by chance alone was also 2%.

However, the practical conclusions are likely to be different. The 2% probability of getting the result by chance would be unlikely to convince people that telepathy is at work because telepathy is generally seen as very unlikely so the "2%" evidence would probably not be sufficient. On the other hand, vegetarianism improving wellbeing seems, to me, plausible so the "2%" evidence seems reasonably convincing.

These percentages are called *p values*. An awful lot of statistical theory is about p values. The hypotheses that p values are based on are conventionally called the null hypotheses - the chance hypothesis in both of our examples - but I prefer the term baseline hypothesis because it is the baseline against which the evidence is assessed.

If the p values are bigger, the conclusions are less useful. With the sample of 3 in each group (the rows of data shown in Figure 1, and resamples like the one shown in Figure 2) the p value is about 30% (just delete the last 54 rows of data in the Sample sheet of the spreadsheet). This suggests that differences as big as +10 or as small as -10 could well happen by chance - for example the shuffle in Figure 2 - so there is no clear evidence either way: it might be chance or there might be a real difference in the population. The *smaller* the p value, the *stronger* the evidence against the baseline hypothesis of no difference between the groups.

The shuffle test is very flexible. You might be interested in the difference between the proportion of old people and young people who vote Labour, or in the correlation between height and intelligence: in each case the shuffle test, and the spreadsheet at http://woodm.myweb.port.ac.uk/SL/resample.xlsx, can be used to work out p values. It does the work of several conventional methods of deriving p values (see below).

## Approach 2: How confident should we be about our conclusions?

The trouble with Approach 1 - testing a baseline hypothesis - is that it doesn't actually tell us how probable our hypotheses are. For the telepathy experiment we discovered there's only a 2% chance of Annette guessing correctly by guesswork alone. The shuffle test told us that there about a 2% chance that similar results could have occurred if there were no difference between veggies and omnies. But how likely is telepathy, and how likely is that veggies overall do have higher wellbeing scores than omnies? We don't know. (If you think you do know, you are wrong[5].) Testing a baseline hypothesis doesn't tell us. Approach 2 does give us tentative estimates of these, more useful, probabilities.

I'll start with an example which is straightforward for the method I'll show you here. Then we'll be in a better position to try the two examples above

### A simple example of bootstrapping

The method used here is called bootstrapping. I'll start with a simple example: the well-being scores of the first 9 vegetarians in the sample summarised in Figure 1 were:

$$74, 65, 57, 78, 54, 47, 38, 34, 93$$

The average of these numbers is 60. Can we assume that this is an accurate measurement of the average well-being of *all* vegetarians? Obviously not because the sample is small and the numbers range from 34 to 93, so chance will mean that the estimate might be too high or too low. But how much too high or low?

For the sake of a definite storyline let's assume that we want to know whether the average is more than 50. The data suggests it probably is, but how probable is "probably"? Remember we're talking about the average here: obviously some measurements are more than 50 and some less.

The idea behind bootstrapping is to do some experiments to see how much the averages of random samples of 9 from this sort of population are likely to vary. The difficulty, of course, is that we've only got data from 9 vegetarians. We don't know about the whole population. The trick we use is to *assume* that this data does give a reasonable impression of what the whole population of vegetarian well-being figures would look like. To be precise, we assume that one ninth of them (11.1%) are 74s, a ninth are 65s, and so on. This is obviously just a guess, so I'll call it a *guessed population*.

Obviously, with very small samples - say 2 or 3 - this is not really viable. Even with a sample of 9 you might feel we are likely to get a restricted view of what the population might be like. This is a fair point, but a sample of 9 is all we've got , and experience suggests we do get reasonable answers.

Now we need to take random samples of 9 from this guessed population. In the physical world, you could do this by writing all nine wellbeing scores on a card, shuffling the pack and dealing one, then *replacing* it so that the pack contains the same nine cards, shuffling and dealing another, and so on nine times. This means that each card dealt is equally likely to be any of the nine scores in the sample of data - which is what we would get if we had a large population with same pattern as the sample. This process is called *resampling with replacement*, to contrast it with the hands dealt in the ordinary way that we used above for the [shuffle test](), which is *resampling without replacement*. Figure 4 shows the original sample and first three resamples I generated in this way.

*Figure 4. Sample of 9 and first three resamples with replacement*

| Sample (Guessed population comprises equal numbers of each member of this sample) | | | | | | | | | |
|---|---|---|---|---|---|---|---|---|---|
| 74 | 65 | 57 | 78 | 54 | 47 | 38 | 34 | 93 | Average=60 |
| First three resamples | | | | | | | | | |
| 93 | 34 | 54 | 93 | 65 | 47 | 78 | 54 | 93 | Average=67.9 |
| 74 | 78 | 34 | 65 | 54 | 93 | 78 | 38 | 78 | Average=65.8 |
| 47 | 74 | 38 | 57 | 78 | 47 | 74 | 47 | 34 | Average=55.1 |

Notice here that, for example, 93 appears three times in the first resample, 54 appears twice and several numbers don't appear at all, so the resample is not the same as the sample. These three resamples do vary quite a lot: from an average of just over 55 to almost 68.

It is, of course, important that the sample should be a random one, or at least like a random sample. This is for two reasons. First it should make it likely that the guessed population will be a similar sort of shape to the real population. Second, the resamples are drawn using a random process which is only likely to be helpful for telling us something about the real samples if they are also drawn using a random process.

There are only three resamples in Fig. 4. Obviously we need more to get a reasonable assessment of how variable these resamples are. To do this we can use the same spreadsheet as we used [above](), but set up with different data and menu settings:
[http://woodm.myweb.port.ac.uk/SL/resample9bootstrap.xlsx](http://woodm.myweb.port.ac.uk/SL/resample9bootstrap.xlsx).  To see how this works click on the Single resample tab, and press F9 (to force the spreadsheet to produce another set of random numbers). Then click on the Lots of resamples tab to see a diagram like Figure 5 which shows the pattern from 1000 resamples generated like this.

*Figure 5. Histogram showing averages of 1000 resamples with replacement from the sample of 9 well-being scores in Figure 4*

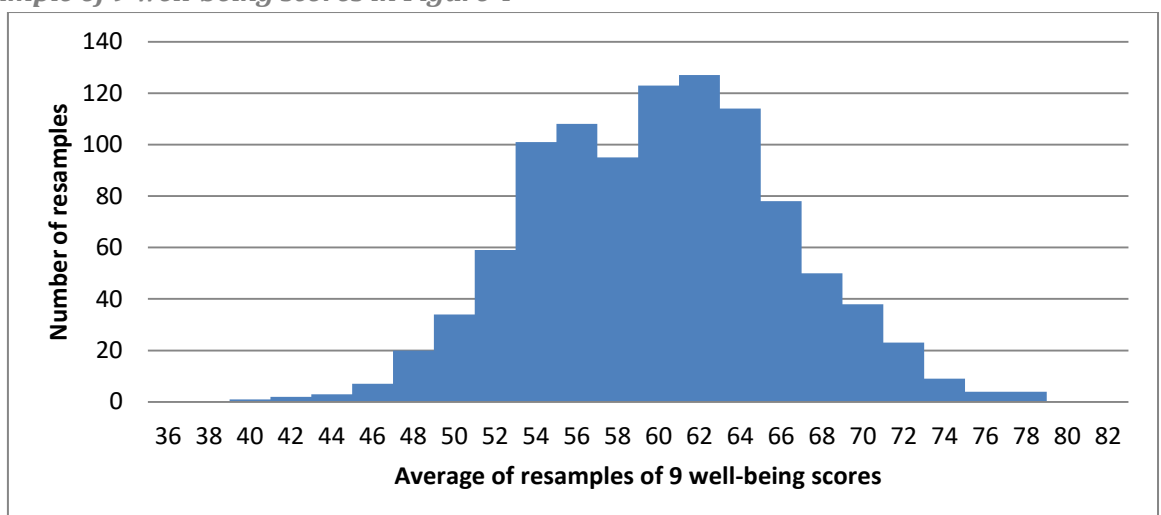

The bar centred on 50, for example, indicates that 34 of the 1000 resample averages lay between 49 and 51. The first resample average in Figure 4 (67.9) is one of the 50 resample averages in the bar centred on 68.

This shows that the averages of these resamples vary a lot: from about 20 units below the overall average (60-20=40) to about 20 units above (60+20=80). *This suggests that plus or minus 20 is the maximum likely error in using a sample of size 9 to tell us about the overall average.*

Now let's use these resample experiments to see what we can say about the average of the real population. We've got a sample of 9 with an average of 60. But we know from the resampling that averages of samples of 9 have an error of +/- 20 if we use them as an indication of the overall average of the population. So the overall average of the real (as opposed to the guessed) population is likely to be within 20 units of the sample average of 60 - i.e. between 40 and 80.

The graph in Figure 5 goes from 20 below the sample average (60-20=40) to 20 above the sample average (60+20=80). This sort of suggests that Figure 5 tells us about our level of confidence in various possible averages for the overall population. This conclusion is represented by Figure 6 which is identical to Figure 5 except that the axes have been relabelled and the vertical axis rescaled to represent probabilities or confidence levels. The "sort of" acknowledges the fact that there are a few potential flaws in this conclusion - see the more detailed justification below if you are interested - although in rough terms it works surprisingly well.

*Figure 6. Confidence levels or probabilities for the population average based on Figure 5*

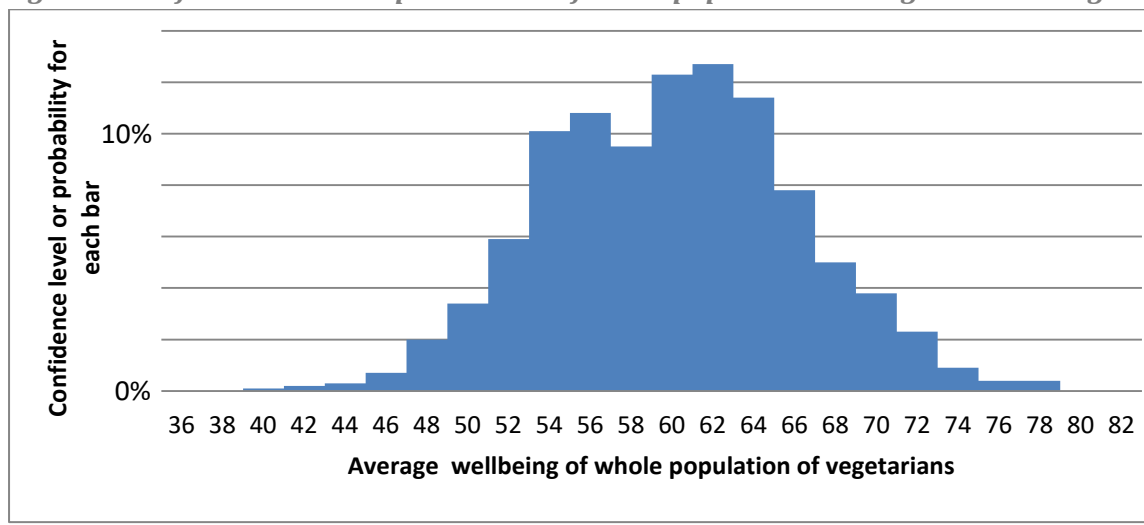

The principle of this is very straightforward. All we've done is guessed that the pattern in the whole population of vegetarians is like the sample, generated lots of random samples from this guessed population, and then assumed that the resulting distribution represents our confidence in various possible population averages. This is known as bootstrapping because it feels as if it's impossible like pulling oneself up by one's bootstraps: we managed to derive conclusions about the whole population from a sample of just 9 with no other inputs except the assumptions that the sample is a random one.

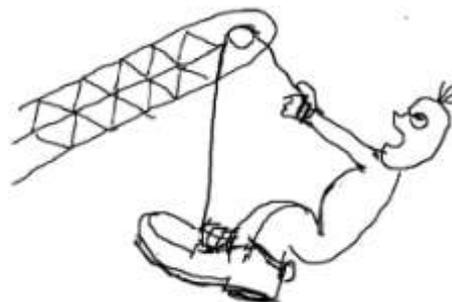

We can now use Figure 6 to answer the question we started with: how confident are we that the average wellbeing of the whole population of vegetarians is 50 or more on the evidence of the sample of 9? The answer is about 95%. It should be roughly obvious that about 95% of the area in the bars in Figure 6 is above 50, and the remaining 5% is below. The exact answer of 95.2% comes from the spreadsheet which simply counts up the number of resample averages which are 50 or more (952 out of 1000). So we can be fairly sure that average wellbeing is 50 or above but not completely sure. Estimates like this are likely to be wrong on about 5% of occasions.

Figure 6 and the spreadsheet it behind can also be used to derive a *confidence interval*: we can be 95% confident that the population average is somewhere between 49 and 72. The estimate from the sample of 60 turns out to have quite a large margin of error.

These two conclusions give us useful probabilities: we have a direct measure of how confident we can be about our conclusions. Approach 1 does not do this. With the 50 card telepathy experiment the probability of Annette getting the card right by chance is 2%. But this is *not* the probability of the chance hypothesis being right - this would imply that the probability of telepathy being the explanation was 98% which most people would consider absurd!

However, despite its usefulness, the logical structure of the bootstrap argument may feel a bit dubious. We start by assuming the guessed population corresponds to the real population, although we are almost sure it doesn't. We then use this - almost certainly false - assumption to estimate how big the error is likely to be. Bootstrapping involves what might be called semi-circular reasoning, which you might feel shouldn't work. But it does! The answers it gives are often surprisingly close to the answers derived from conventional probability theory - see [below](below).

## Can we apply bootstrapping or Approach 2 to the telepathy experiment?

It would obviously be nice if we could apply approach 2 to the [telepathy experiment](telepathy experiment) to see what the probability of telepathy was. But this is not possible. If it were we would get a probability, like the 95% above, representing the probability of telepathy. Now imagine someone who believes firmly that telepathy is completely impossible. Whatever the probability of a correct guess, she would say that as telepathy is impossible, chance *must be* the correct explanation. On the other hand someone who was more open minded might be swayed by the million card experiment but perhaps not by the 50 card experiment.

This means that any reasonable answer must take into account how plausible we think the hypothesis is, so there cannot be a general method which ignores this. Besides this, the example of Approach 2 above starts with a range of possible values for what we are trying to estimate (the horizontal axis in Figure 6). We don't have this in the telepathy experiment.

Approach 2, then, can't be used to analyse the telepathy experiment. But there is an extension of Approach 2, which takes account of so called prior probabilities, which can be used. This is based on a theorem invented by the Reverend Thomas Bayes published posthumously in 1763 which has spawned an approach to statistics called Bayesian statistics - see [below](below).

## Bootstrapping the wellbeing of vegetarians versus omnivores

On the other hand, there is no difficulty applying Approach 2 to the difference between the vegetarians and omnivores. (This is the question we tackled [above](above).) We start with the sample of vegetarians and omnivores and compare their well-being as in [Figure 1](Figure 1). The veggie's average well-being is 10 units higher than the omnivores' in this sample of 60: the difference, average veggie wellbeing - average omnivore wellbeing, is 10.

We can now just use the bootstrap trick (see [above](above)) to see how accurate this is likely to be as an estimate of the *difference* in average wellbeing of the whole populations of veggies and omnivores. This set up on the spreadsheet at
[http://woodm.myweb.port.ac.uk/SL/resample60bootstrapdifference.xlsx](http://woodm.myweb.port.ac.uk/SL/resample60bootstrapdifference.xlsx).

We start by choosing one of the 60 rows of data (see [Figure 1](Figure 1)) at random, then replace it and choose another one, and so on till we have a resample of 60. When I did this the resample had 32 veggies with an average score of 62.8 and 28 omnivores with average score of 44.6 giving a difference of 18.2. (To see how this works click on the Single resample tab in the spreadsheet.) The spreadsheet then produces 1000 resamples in the same way giving the pattern shown in Figure 7 (click on the Lots of resamples tab to see this).

There are two differences between this and process used for [Figure 3](Figure 3): bootstrapping uses resampling *with* replacement, and the random choice is a whole row of data - a wellbeing score and the information about whether it comes from veggie or omnivore so that these are kept together - instead of shuffling to allocate the wellbeing scores to veggies or omnivores at random. The same spreadsheet is used for both with different menu options chosen in the brown cells.

*Figure 7. Histogram showing the average difference between veggies and omnivores in 1000 resamples with replacement from the data in Figure 1*

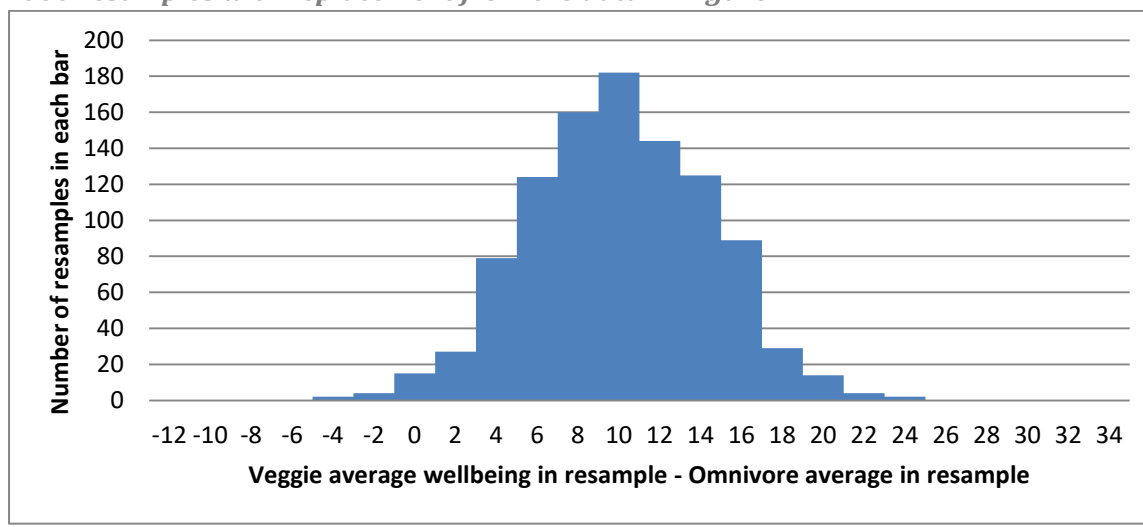

Just as in the example above it is reasonable to interpret this as telling us about the probability of various possible differences between the veggie average and the omnivore average in the whole population. This suggests it can vary from about -4 (omnivore average is 4 units more than veggie) to about +24 (veggies 24 units higher). The detailed statistics for the difference in the averages from the spreadsheet tell us that the probability of the difference being more than zero - in other words of the veggies having a higher average wellbeing than the omnivores - is 98.8% or about 99%. Which is what we wanted to know: vegetarianism is almost certainly associated with higher wellbeing.

Another question we could answer is about how likely it is that veggies will score *substantially more* than omnivores. If substantially means more than 10 units the answer is obviously about 50%, if it means more than 5 units the answer is about 87%. And so on.

Yet another conclusion is that the 95% confidence interval for the difference in average wellbeing extends from 1.2 to 18.3. We can 95% sure that the true average lies somewhere in this interval. This gives us an idea of the likely error in the estimate from a limited sample.

This method is surprisingly flexible. It can, for example, be used to estimate confidence levels for various election outcomes from opinion poll data (see below), and to see how accurate a correlation based on a sample is likely to be (see Wood, 2004 or Wood, 2005, but note that the spreadsheet at http://woodm.myweb.port.ac.uk/SL/resample.xlsx is an improved version of the software used in these articles).

## General summary and more details of Approaches 1 and 2
I've explained Approaches 1 and 2 by means of examples. It's now time to summarise each in general terms. This is important because the underlying ideas are far broader than the examples and methods above.

## Approach 1 is testing a baseline hypothesis

... by estimating how probable the data, or similarly or more extreme data, is on the assumption this hypothesis is true. The baseline hypothesis is often called a null hypothesis, and the general approach null hypothesis significance testing or NHST (see below).

The resulting probability is called a p value. Small p values indicate that the data is unlikely to have arisen if the baseline hypothesis is true, so this hypothesis is unlikely to be true.

This is a sort of fuzzy, statistical version of reductio ad absurdum which is a commonly used approach in maths. For example, to show there is no biggest whole number, assume there was, add one to it and you have a bigger whole number, which contradicts the original assumption - so the original assumption that there is a biggest whole number must be wrong. Similarly to do a statistical test, you assume the baseline hypothesis is true, estimate how likely your data is to have arisen on this assumption, and if this probability is small you conclude the baseline hypothesis is likely to be false.

How we work out the p value depends on the circumstances. For the telepathy story it's pretty obvious how to do it. The method used for the comparison between the wellbeing of vegetarians and omnivores was the shuffle test. And there are dozens of other possibilities in statistics textbooks (see below).

## Approach 2 is estimating tentative probabilities of whatever we are interested in being true.

Again there are many possible ways of doing this, although sometimes, as with the telepathy story, it may not really be possible. Bootstrapping is a useful and very flexible approach. Conventional statistics only uses Approach 2 in relation to confidence intervals - and there are many established ways of deriving these which can be adapted to estimate confidence levels or probabilities of more general hypotheses as explained below.

Bayes theorem is the rigorous mathematical underpinning of Approach 2, but so-called Bayesian statistics is a small sect outside mainstream statistics. Bayesian methods tend to be on the complicated side, although the principle is relatively straightforward.

What both approaches have in common is that we are considering a range of possibilities with a view to deciding which is most likely to true. The starting point is always making a list of these possibilities: telepathy or guesswork in the telepathy story, and in the second story the difference between the veggie's and the omnivores' average wellbeing (rounded to the nearest 2) being -6, -4, -2, 0, + 2 or any of the other values on the horizontal axis of Figure 7.

Where the approaches differ is in what they do with these possibilities. Approach 1 starts with one of these possibilities as a baseline (guesswork in the telepathy story, and a difference of zero in the veggie-omnivore comparison), and then assumes this baseline to work out the probability of getting the data or more extreme data (the p value). Approach 2, on the other hand, starts with the data, and then uses this to estimate a probability for each possibility (Fig. 7 for the veggie-omnivore comparison).

These probabilities are the reverse of the p values in the same way that if you *start with a with all the picture cards in a pack*, the probability that a randomly chosen one is a king is 1/3 (because there are 4 kings in the 12 picture cards), whereas *if you start with all the kings*, the probability that a randomly chosen one is a picture card is 100% (because all four kings are picture cards). These two probabilities are, of course, very different. The probability of Annette choosing the correct card by guessing in the [telepathy experiment](#) is 2%, but the probability of Annette being telepathic given that she's chosen the correct card is a very different probability (most people would think much less than 2%).

Approach 2 is generally more useful. You get a direct probability for whatever is of interest, instead of a convoluted, reversed probability based on a hypothetical baseline hypothesis (like the no-difference hypothesis in the veggie-omnivore comparison). However, the downside is that the probabilities for Approach 2 must be somewhat *tentative*. You need to make assumptions which may not be reasonable. Is it, for example, reasonable to assume that the guessed population constructed from the sample in the bootstrap procedure will give us reliable estimates of how much sample averages will vary? Obviously we can never be sure. And in the telepathy story, the fact that a reasonable answer must depend on prior beliefs means that there can be no single answer that everyone will accept, so any probabilities derived must be tentative.

On the other hand for Approach 1 the probabilities are likely to be much less tentative. It is difficult to quibble with either the 2% p value for the telepathy story, or the 2% p value for the comparison between vegetarians and omnivores, because they are both derived from straightforward arguments involving equally likely possibilities, whereas the bootstrap method has far more potential flaws (see [below](#)) so the results must be regarded as more tentative.

Approach 1, then, leads to definite results which may not be very useful. Approach 2 leads to results which are tentative but probably more useful.

### Both approaches: the important small print

The purpose of this sort of statistical analysis is usually to make inferences from a sample which go beyond the actual data gathered. In the telepathy story the aim is to infer a likely explanation for the observed fact that Annette has chosen the right card. In the veggie-omnivore comparison the idea is to generalise the results to *all* veggies and omnivores in the population. Which, of course, begs the question of what the population is - everyone in the world, or in a particular country? Does it include those who might become vegetarians in the future? The phrase *target population* is sometimes (helpfully) used to emphasise the idea that the population is whatever the research is targeting[6].

The sample, and how it was chosen, is crucial. The sample should obviously be typical, or representative, of the wider context of interest. If there is a definite population, the best way of doing this is often to choose a sample at random from the wider population. If you have a list of this wider population, this can easily be done with random numbers[7].

In practice random samples are often difficult to obtain. You obviously can't have a list of future vegetarians! And biases can easily creep in. Some people may refuse to answer your questions, or be unavailable, so your sample will be skewed in the direction of those willing and able to answer - who may have very different views from those you don't manage to speak to. If you are doing a phone survey and ask the first person to answer the phone, your sample will be biased in favour of people who like answering the phone, and in favour of those in small households (if you live by yourself you will have a 100% chance of being the one answering the phone, whereas if you live with nine other people your chance will be around 10%). It is important to be aware of these biases and do your best to deal with them.

There is also an important distinction between (1) a survey of people where you take a random sample of people and ask them if they are vegetarians, and (2) a survey where you take a random sample of vegetarians and another sample of omnivores, and (3) a trial or experiment where you start with a group of omnivores and then tell a randomly chosen group of them to go vegetarian so that you can see the impact this has on their wellbeing. (3) means that any difference in wellbeing you observe is likely to be caused by going vegetarian, whereas (1) or (2) don't enable you distinguish cause and effect: it may be that people with a higher wellbeing score tend to prefer vegetarian food so the causal link may be in the opposite direction.

These points are vitally important. And it's also important to remember that measures like the wellbeing test might be suspect, the definition of a vegetarian may be a bit hazy, potential voters in an opinion poll might lie or change their minds, and so on. The answers you get from the statistical methods like the shuffle test or bootstrapping are likely to underestimate the degree of uncertainty, simply because all these other factors are ignored.

## Is there only one right answer?

No. In two rather different senses as explained in the two subsections below.

### Random methods and random answers: simulation methods versus probability calculations

Most of the methods above rely on computer simulation and random numbers, which means that every time you run them the answer will be slightly different. For example, I've just pressed F9 six times to generate six p values for the [shuffle test](#) explained above, and the answers were:

$$2.4\%, 2.6\%, 2.0\%, 2.3\%, 2.8\%, 2.0\%$$

Each of these is based on 1000 resamples and, not surprisingly, they vary a bit so one answer will be slightly different from the next one.

On the other hand the analysis of the telepathy experiment yielded a single p value: 1/50. No simulation, so you get the same answer every time. This is a very simple example of the use of probability theory; most statistical theory is based on probability theory and so each method gives a single definite answer.

Does the fact that you get a slightly different answer each time you use a simulation method matter? Yes, because if there is one right answer most of them must be wrong. More seriously, a lot of researchers are keen to get their p values below a certain threshold because this is taken to indicate that their data is "significant" in the sense that it signifies a genuine effect and not a chance fluctuation: if 2% was such a threshold then it would be tempting to carry on pressing F9 until you get a result below 2% and then take that as the "answer". Which is really cheating and a distortion of the true probabilities. In an ideal world, everyone would take the first result obtained. (The same problem occurs in conventional research using probability theory where each set of data yields just one p value. The problem here is that if the p value isn't satisfactory, it is tempting to get another set of data, or use a different statistical test, in the hope that you will get a "better" p value.)

On the other hand the fact that p values calculated with probability theory are always the same with the same data may give a false sense of security. Another sample would give another p value, so the stability of the answer is an illusion which simulation methods may dispel.

Of course, the variability of the p values can be reduced by simply taking more resamples. Taking all 6000 resamples used for the six p values above, the result (the average) is 2.35. If you took another 6000 the resulting p value would not be identical, but it would be closer than the values above based on 1000 resamples. By taking more resamples you can make the answer as accurate as you want.

**Tentative versus definite probabilities**
There is an important difference between the 2% p value for the shuffle test, and the 95% probability for the average wellbeing of all the vegetarians being more than 50 which we worked out by bootstrapping.

The rationale behind the shuffle test is almost beyond dispute: we want to know how likely a difference of 10 between the vegetarian and the omnivore averages is if it's all a matter of chance, so we simply do lots of shuffles - which is what we mean by chance - and see how many have a difference of 10 or more. This seems definite in the sense that no reasonable person could argue with the rationale. (Or, at least that's the way it seems to me: with probabilities most arguments have a counter-argument!)

The bootstrap argument is much more tentative. The argument above for justifying the bootstrap methods includes the phrase "sort of" to indicate that it is a bit suspect. I've gone through some of the assumptions on which the validity of the bootstrap method depends [below](). It would be quite reasonable to say, for example, that the "prior probabilities" of each possible difference may not be equal. Most conventional statisticians would not even accept the 95% as a genuine probability: at best they would call it a confidence level[8]. But it is definitely far more tentative than the 2% from the shuffle test. Which, means, of course, that we shouldn't treat the 95% as being the single right answer in the same way as it would be reasonable to assume this of a shuffle test probability (provided we used enough resamples).

## Sample sizes
Have you ever been asked about your voting intentions by an opinion poll? I haven't, and I suspect most people haven't either. But the opinion polls still claim to know how everyone, including me, will vote *despite the fact that they haven't asked me*! They don't get the answers dead right, but they are usually in the right ball park. This is the magic of statistics: achieving what, at first sight, shouldn't be possible.

It's easy to do an experiment to see how this is possible, and how small a sample we can use and still get useful answers. Let's imagine we've got 500 voters in the electorate all of whom voted, and 300 of them (60%) voted for Socrates. An opinion poll before the election was based on a sample of just 20 of them.

I've set this up in the spreadsheet [http://woodm.myweb.port.ac.uk/SL/resamplepoll.xlsx](http://woodm.myweb.port.ac.uk/SL/resamplepoll.xlsx). If you click on the Sample tab at the bottom you will see the 300 votes for Socrates (the ones in the green column) and the 200 votes for other candidates (coded as zeros which you will see if you scroll down). (The use of the word "sample" comes from the original purpose of the spreadsheet - click on the Introduction tab for details.) If you click on the Single resample tab you will see a random sample of 20 people from this sample. When I did this there were 9 ones corresponding to 9 votes for Socrates and 11 zeros corresponding to 11 votes against him giving a 45% vote for Socrates (0.45 in the spreadsheet which just takes the average of 9 1s and 11 0s). When I pressed F9 to recalculate the random numbers I got 50%, and when I did it again 65%. Each of these corresponds to a different random sample of 20 electors.

Is a sample of 20 accurate enough to be useful? The Lots of resamples worksheet gives the results of 1000 resamples, corresponding to 1000 polls of 20 electors: the results vary from 30% (0.3) to 85% with 95% between 40% and 80% (the 2.5 and 97.5 percentiles). Which doesn't really feel good enough. If the poll result were 40% when the actual result is 60% this is a serious error.

So let's make the sample bigger, say 100. When I changed the size of the resample (poll) in the Single resample sheet, the 95% interval went from 51% to 68%. Better, but perhaps not good enough. Obviously you could experiment with bigger sample sizes to see what effect this has.

There is another point to consider here. What if the number of electors were a lot larger - say millions of them? Intuitively, you might think that this would make the poll a lot less accurate, but you'd be wrong: it makes surprisingly little difference.

We can simulate this on the spreadsheet by resampling *with replacement*. This is achieved by changing the menu option in the brown cell on the left hand side of the Single resample sheet. Every time an elector is chosen at random for the sample, it is replaced so that each random selection is from the same group with 60% Socrates voters. This is just what would happen if there were millions of voters (or an infinite number of them): removing one will make no practical difference so there will always be a 60% chance of choosing a Socrates voter.

When I did this the 95% interval for the poll with 20 electors was 40% to 80% (just as before), and for the poll with 100 electors was 51% to 69% (instead of 51% to 68%). Having an infinite or very large population seems to make no difference for the small poll, and only very slight difference for the larger poll.

It is actually fairly obvious why this should be so. The only difference between the two sampling methods is that if we sample without replacement the pool of electors left to sample will be slightly different depending on which electors have already been sampled. But, unless the size of the sample is close to the whole population of electors, this is only going to make a very small difference.

This means that a sample of 100 electors will give almost as accurate an assessment of a population of 500 as it will of a population of 500 million electors. The first sample involves polling 20% of the electorate, and the second only 0.000002%! In other words, it is irrelevant what percentage of the population you are sampling; what matters is the number - 100 in this case.

The typical sample used by polling companies in UK elections is 1000. Let's assume again that Socrates is a candidate, and that our initial poll of 500 suggests he's got 60% of the votes. We can now use the spreadsheet to resample with replacement to see how effective a sample of 1000 electors would be - simply be setting the resample size to 1000. (This necessitates increasing the maximum resample size to 1000 as explained in the Extras tab of the spreadsheet.) The result is that the 95% interval goes from 57% to 63% implying an error of 3% in either direction. Which is obviously much better than the smaller samples.

It is easy to use a similar approach to investigate the effect of changing the sample sizes in all the examples above. For example, the shuffle test above produced a p value of 2% - reasonably convincing evidence. What sort of evidence would a bigger sample have produced? All we need to do to answer this question is to copy the data in the sample sheet and paste it below the data (starting in Cell D65) so that we now have twice as much data but with the same difference between veggies and omnivores. Now go to the Lots of resamples sheet and press F9 to recalculate the spreadsheet. When I did this the histogram was between -10 and +10 (instead of -14 and +14) and the p value was 0.1% (instead of about 2%).

So the bigger sample produces a much lower p value corresponding to stronger evidence. As we would expect. What about a smaller sample? When I deleted half the sample, the p value increased to 8.3% corresponding to weaker evidence - again as we should expect. The conventional cut-off is 5%, so this would be considered too small a sample. The original sample of 60 seems about right.

## Bayes' Theorem via the Deleted Worlds Story

In the telepathy experiment above Annette chose the right card from the pack of 50 cards. There were only two hypotheses to explain this: either she was guessing or telepathy was involved. Bayes' theorem gives us a way of deciding which is more likely.

We need to start with how likely we think the two hypotheses are because a reasonable answer for someone who thinks telepathy is plausible must be different from a reasonable answer for someone thinks telepathy is impossible. For someone who thinks telepathy is impossible the probability of telepathy would be zero, and guessing, however unlikely it would be to deliver the right answer, is the only viable alternative. On the other hand, for someone who thinks telepathy is a plausible hypothesis[9], our experiment would obviously make it more plausible.

Let's suppose we think - before doing the experiment - that guessing is three times as likely as telepathy, so the *prior probability* of guessing is 3/4 and of telepathy is 1/4.

The intuitive basis of probability is games of chance in which probabilities are the proportion of equally likely possible outcomes which have the characteristic of interest. The probability of dealing a spade from a pack of 52 cards is 13/52 or 1/4, for example, because 13 of the 52 cards are spades. We can we extend this idea to our problem here by imagining lots of equally likely possible worlds, in some (a quarter) of which Annette is guessing, and in the others (three quarters) she is using telepathy. (There's a slightly more extended discussion of this in the Appendix.) Figure 8 shows 200 such possible worlds.

*Figure 8: Two hundred possible worlds*

| | |
|---|---|
| 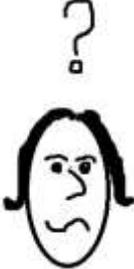 | 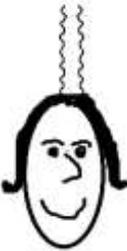 |
| *150 **G**uessing Worlds* | *50 **T**elepathy Worlds* |
| G✓ G✓ G✓ G✘ G✘ G✘ G✘ G✘ G✘ G✘ G✘ G✘ G✘ G✘ G✘ | T✓ T✓ T✓ T✓ T✓ |
| G✘ G✘ G✘ G✘ G✘ G✘ G✘ G✘ G✘ G✘ G✘ G✘ G✘ G✘ G✘ | T✓ T✓ T✓ T✓ T✓ |
| G✘ G✘ G✘ G✘ G✘ G✘ G✘ G✘ G✘ G✘ G✘ G✘ G✘ G✘ G✘ | T✓ T✓ T✓ T✓ T✓ |
| G✘ G✘ G✘ G✘ G✘ G✘ G✘ G✘ G✘ G✘ G✘ G✘ G✘ G✘ G✘ | T✓ T✓ T✓ T✓ T✓ |
| G✘ G✘ G✘ G✘ G✘ G✘ G✘ G✘ G✘ G✘ G✘ G✘ G✘ G✘ G✘ | T✓ T✓ T✓ T✓ T✓ |
| G✘ G✘ G✘ G✘ G✘ G✘ G✘ G✘ G✘ G✘ G✘ G✘ G✘ G✘ G✘ | T✓ T✓ T✓ T✓ T✓ |
| G✘ G✘ G✘ G✘ G✘ G✘ G✘ G✘ G✘ G✘ G✘ G✘ G✘ G✘ G✘ | T✓ T✓ T✓ T✓ T✓ |
| G✘ G✘ G✘ G✘ G✘ G✘ G✘ G✘ G✘ G✘ G✘ G✘ G✘ G✘ G✘ | T✓ T✓ T✓ T✓ T✓ |
| G✘ G✘ G✘ G✘ G✘ G✘ G✘ G✘ G✘ G✘ G✘ G✘ G✘ G✘ G✘ | T✓ T✓ T✓ T✓ T✓ |
| G✘ G✘ G✘ G✘ G✘ G✘ G✘ G✘ G✘ G✘ G✘ G✘ G✘ G✘ G✘ | T✓ T✓ T✓ T✓ T✓ |

✓ means correct card chosen; ✘ means incorrect card chosen

Figure 8 also shows whether Annette chose the right card in each of these worlds. If she's guessing there is a one in 50 chance of getting it right, so she'll get it right in three of the 150 guessing worlds. These are indicated by a tick in Fig. 8. Similarly if she's telepathic, she'll get the right answer in all 50 of the Telepathy worlds (we're assuming telepathy is 100% accurate). Now all we have to do is remember that because she did guess correctly the real world is one of those with a tick. *We can delete all of 147 worlds in Figure 8 without a tick because these are not consistent with the results of the experiment.* There are 53 of these, of which 3 are telepathy worlds - as in Figure 9.

*Figure 9: The 53 remaining possible worlds after the 147 worlds that are not consistent with the data have been deleted*

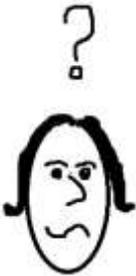

| 3 **G**uessing Worlds | 50 **T**elepathy Worlds |
|---|---|
| G✓ G✓ G✓ | T✓ T✓ T✓ T✓ T✓<br>T✓ T✓ T✓ T✓ T✓<br>T✓ T✓ T✓ T✓ T✓<br>T✓ T✓ T✓ T✓ T✓<br>T✓ T✓ T✓ T✓ T✓<br>T✓ T✓ T✓ T✓ T✓<br>T✓ T✓ T✓ T✓ T✓<br>T✓ T✓ T✓ T✓ T✓<br>T✓ T✓ T✓ T✓ T✓<br>T✓ T✓ T✓ T✓ T✓ |

As all of these 53 worlds are equally likely (we started by assuming we had 200 equally likely worlds so those remaining will still be equally likely), we can work out probabilities just like a game of cards. The probability of telepathy is 3/53 or 5.66%. And the probability of her guessing is 50/53 or 94%.

These are often called *posterior* probabilities to distinguish them from prior probabilities. The evidence of the experiment has increased our assessment of the probability of the telepathy hypothesis from 25% to 94%. Problem solved!

This is just a model or a story[10]: something we imagine to help us solve a problem. I'm not saying these worlds are real (unlike the proponents of the many worlds interpretation of quantum mechanics), just that they provide a helpful way to make sense of the problem. Why did I choose 200 worlds? Because this gives me a multiple of 50 guessing worlds which is helpful for the next step. I could have chosen 400 or 1000, but 100 would not work. You may be thinking that there won't be *exactly* 3 of the 150 where she gets the right card, but on average there will be, and this is (obviously?) all we need to work out the probabilities.

This story of possible worlds mirrors the formulae of the mathematical version of Bayes' theorem[11]: it isn't an approximation for the unintelligent, but an intuitive model for those who want a full understanding as quickly as possible.

The story in Figures 8 and 9 can be modified in two obvious ways:

*We can start with different prior probabilities.* If, say, we thought guessing and telepathy were equally likely (both prior probabilities 50%), then we might have 50 worlds in each camp and the probability of telepathy would be 50/51 or 98%. If we thought the prior probability for telepathy was only 1%, we might have 50 telepathy worlds and 4950 (99*50) guessing worlds and the probability of telepathy would be 50/149 or 34%. If you think the prior probability of telepathy is zero, there would be no telepathy worlds in our model and the probability of telepathy would be zero. *Bayes theorem gives us a way of adjusting the prior probabilities to take account of the evidence.*

*What if the success rate of telepathy is less than 100% - say 50%*. Then half of the 50 telepathy worlds in Figure 8 (25 of them) would be have ticks and the rest (25) would be deleted, so 25 of the 28 tick worlds would be telepathy ones and the probability of telepathy would be 25/28 or 89%.

The general scenario for Bayes' theorem is that we've got a list of possibilities which don't overlap and of which one, and only one, is true or right. The difficulty is that we'd like to know the probability of each possibility being true, but there is no easy way of doing this. However, if we have some data, or something which we've observed, and we can work out the probability of this data or observation having occurred *if* each of the possibilities are true, then we can use Bayes theorem. (In Fig. 8 the data or observation is that Annette has got the card right, and it's easy to see that *if* she's guessing this would be a 2% chance, whereas if she's telepathic it's a 100% chance.)

To use an equally likely possible worlds model - which lets us plug easily into intuitive ideas of probability - we need to start by deciding how probable each of the possibilities are, and expressing this in terms of equally likely possible world as in Fig. 8. Then for each possibility we work out how many of the equally likely possible worlds are consistent with the data or observations, delete the rest, and then use the remaining worlds to estimate how likely each of the possibilities are (Fig. 9).

Bayes' theorem is for reversing probabilities[12]. For example, it's easy to work out that *if* Annette was guessing, the probability of her choosing the right card is 2%. Bayes' theorem enables us to reverse this is conclude that *if* she chose the right card, the probability of her using guesswork was 50/53.

Bayes theorem has many uses. One scenario where it can be very useful is in deducing the implications of tests for nasty diseases. Many tests for diseases like cancers don't always give the right answer. False positives (you're told you have the disease when you haven't) and false negatives (you're told you don't have the disease although you do) are a common problem.

Let's imagine we've got a test where the rate of false positives and false negatives are both 5%. So if you have the disease there is still a 5% chance the test will say you're OK, and if you haven't got the disease it'll say you have with a 5% probability. Let's suppose you get the test result, and it's positive. How likely is it that you've got the disease? You may think that a plausible answer is 95%, which would be very depressing. But this wrong, very wrong!

The problem is that the 5% probabilities are the wrong way round. *If* you haven't got the disease, the probability of a positive test result is 5%. What you want to know is *if* you get a positive test result, what's the probability of your not having the disease. These two probabilities are not the same and you need Bayes' theorem to work out the second from the first.

The two possibilities are that you have the disease or you don't. Bayes' theorem needs us to start with the prior probabilities of these. The obvious way to estimate these is to use the rates of the disease among the general population. Most diseases are rare (almost by definition): let's say this one has a prevalence of 1%. (If you have other reasons to think you may have the disease you may want to make this a bit higher, but 1% is the obvious starting point.) Then the probability of your not having the disease is 99%. This is the bit which people tend to forget: they are concentrating so hard on the test that they forget to take into account how rare most diseases are.

Now let's imagine 20 worlds in which you have the disease, and 1980 (99*20) in which you don't. remembering the 5% error rate, you'll get a positive test result in 19 of the disease worlds, and 99 (5% of 1980) of the healthy worlds. This makes a total of 118 worlds with a positive test result, but in only 19 of these do you actually have the disease. So the probability of your actually having the disease is 19/118 or only 16%. So cheer up: you're probably OK despite having a positive test result!

And there are many more examples of Bayes' theorem in action. It is widely seen as the paradigm of rational thought. You start with a an initial idea - reflected by the prior probabilities. Then you get some evidence and apply Bayes theorem and work out your posterior probabilities. These are then your best guess; they might be the prior probabilities for another round of Bayes theorem. In this way you can update the probabilities as you get more evidence. For example imagine that Annette did the telepathy experiment again and got the card right a second time. The prior probability of telepathy for the second round is 50/53, and Bayes theorem can be used to update this to get the new posterior probability of 2500/2503 or 99.88%[13]. This is the statistical approach to learning from the evidence.

it's worth putting Bayesian reasoning into a more general framework. Almost all useful thought involves envisaging possibilities[14] which may or may not happen or which may or may not be right: for example planning to prevent a disaster which we hope won't happen, or deciding between various hypotheses, or choosing between different courses of action, only one of which will actually be carried out. The only difference when using statistics is that we have to list the possibilities systematically, and, if we want to make things simple, create the possibilities so that they are all deemed equally likely.

## Bootstrapping: a more careful justification from a Bayesian perspective

I've explained how bootstrapping works and given a fairly rough justification above, and also how Bayes theorem works. I now want to bring these together to show you how Bayes theorem can be used to give a more thorough justification of bootstrapping, and to clarify the assumptions on which the method depends. This bit is a little bit more involved than the rest of this document, and as the conclusion is that the method and rationale above are usually OK provided that the assumptions listed here hold, you could skip this section if you aren't interested in being pedantic.

Perhaps the most obvious assumption in the bootstrap method above is one that really doesn't matter and is not a hole in the argument at all: the fact that we've rounded off all the numbers to the nearest even number in Figure 5. If we rounded them off to the nearest unit, or tenth, the argument above would still apply. In fact, whatever degree of approximation we make, the argument that the resampling distribution represents an estimated probability distribution for the target population still stands.

More important are the assumptions (1) *that the sample is a random one*, or at least reasonably like a random one, and (2) *that the sample is large enough to give us a reasonable idea of the population*. The resamples are chosen at random, so these are only likely to give us a good idea of what real sample might look like if these real samples are also random. And a sample of one or two would obviously be a waste of time, but is 9 big enough? There is no neat answer to this question, although experience suggests that 9 is often big enough to be useful.

To use Bayes theorem to analyse the bootstrap idea further we need to decide what the possibilities are and how likely we consider each to be before getting the sample of data. Let's assume that the possible averages for the whole population are 0, 2, 4, ...58, 60, 62 ... 100 as in Figure. 5. If we are measuring the average to the nearest even number, and 0 and 100 are the top and bottom of the scale, then these are the 51 possibilities.

How likely are these 51 possibilities before getting the sample of data? You might think that 0 and 100 are very unlikely, whereas numbers in the middle like 50 or 60 would be more likely. It's perfectly possible to build assumptions like this in, the but the easy, lazy *Assumption 3 is that all possibilities are equally likely.* Easy, lazy assumptions are often a good bet, as long as we remember they may not be right.

As the resampling spreadsheet works with 1000 resamples the obvious thing to is to have 1000 possible worlds for each of these 51 possibilities, making 51000 possible worlds in all. All equally likely.

The next step is to delete the possible worlds which are not consistent with the data: our random sample of 9 vegetarian wellbeing scores. This is where we need to think about what we can infer from the resampling results in Figure. 5.

Let's start with the bar centred on 60: there were 123 of the 1000 resample averages (12.3%) in this bar. The average of the guessed population (based on the sample) from which the resamples are taken is 60, so this suggests that, if the population average really were 60, and we took a 1000 random samples of 9, about 123 would have an average of 60 like the real sample. Thinking of each sample as a possible world, the other 877 should be deleted, leaving 123 possible worlds corresponding the possibility that the population average really is 60.

Now let's take the bar centred on 58. There were 95 resamples (9.5%) in this bar, which is 2 units *below* the guessed population average of 60. This suggests that the probability of a sample average being 2 units below the population average is 9.5%. The actual sample average is 60, so if this is 2 *below* the population average, the population average must be 2 *above* 60, or 62. So a reasonable inference from this bar is that if we start with 1000 possible worlds with a population average of 62, about 95 of them (9.5%) will be have the actual sample average of 60, so the other 905 can be deleted.

A similar argument applies to all the other bars. The result is that after deleting all the worlds which aren't consistent with the data, we are left with the possible worlds shown in Figure 10. The number of possible worlds corresponding to a population average of 60 is 123, and of 58 is 95. Comparing this with Fig. 5, the 95 bar is opposite 62, and the bar corresponding to 58 in Fig. 10 is the same as the bar corresponding to 62 in Fig. 5. In other words the left side of the Fig. 5 is like the right side of Fig. 10, and the right side of Fig. 5 becomes the left side of Fig. 10.

*Figure 10. Possible worlds consistent with the data*

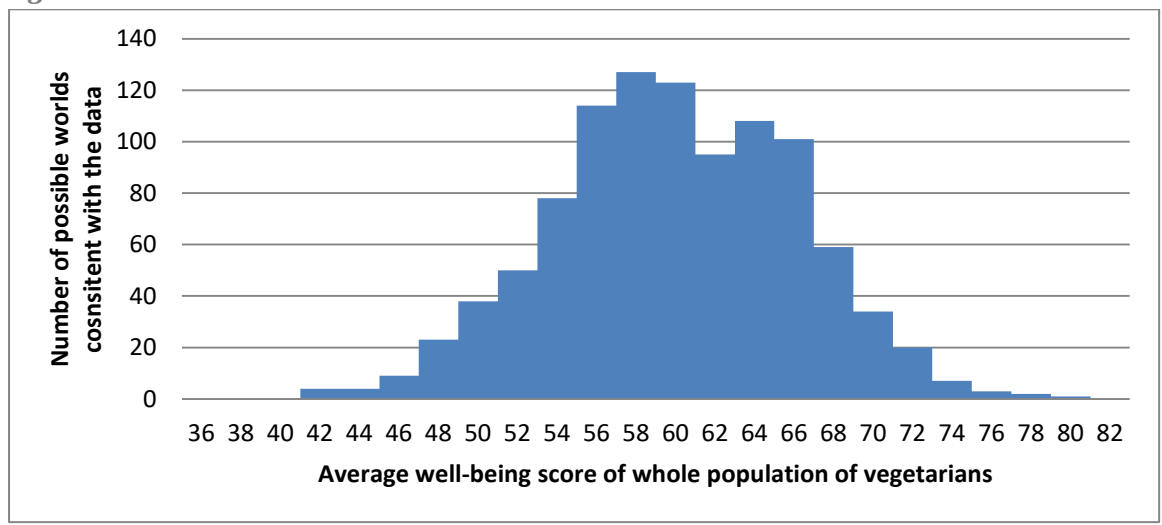

Fig. 10 could be used to read off our confidence level for each bar, remembering that the total number of (equally likely) possible worlds shown is 1000. However, in practice, *if (Assumption 4) the graph is roughly symmetrical* the two sides will be similar so we can ignore this reversal. (The apparent differences between the left and right sides of Figs. 5 and 10 are largely random fluctuations as you will see if you press F9 in the spreadsheet to generate another set of random numbers.)

There is another assumption in the argument above. The bar centred on 80 in Fig. 10 is a prediction about what would happen if the population average really were 80, but this is based on a guessed population with an average of 60. *This assumes (Assumption 5) that the pattern for the guessed population will apply to different population averages*: that the distribution graph can be slid along the axis. In practice, if the graph is not symmetrical this may seem unlikely - there is an example below. For this reason, there is little point in reversing the graph like Fig. 10, because where this is necessary, Assumption 5 is not likely to hold.

To see why these assumptions matter, and the other things that might go wrong with the bootstrap explanation above, I've constructed another sample of wellbeing data, shown in Fig. 11 with the same sort of analysis as before based on the spreadsheet at http://woodm.myweb.port.ac.uk/SL/resample9skewedbootstrap.xlsx, which is designed to be as awkward as possible.

This is another sample of vegetarians, but not a random sample this time, but instead volunteers put forward by a society which advocates vegetarianism. Not surprisingly this sample shows much higher levels of wellbeing and is obviously a strongly biased. Any attempt to infer anything about the population of vegetarians in general from this sample is obviously going to lead to seriously misleading conclusions.

*Figure 11. Histogram showing averages of 1000 resamples with replacement from a second sample of 9 well-being scores*

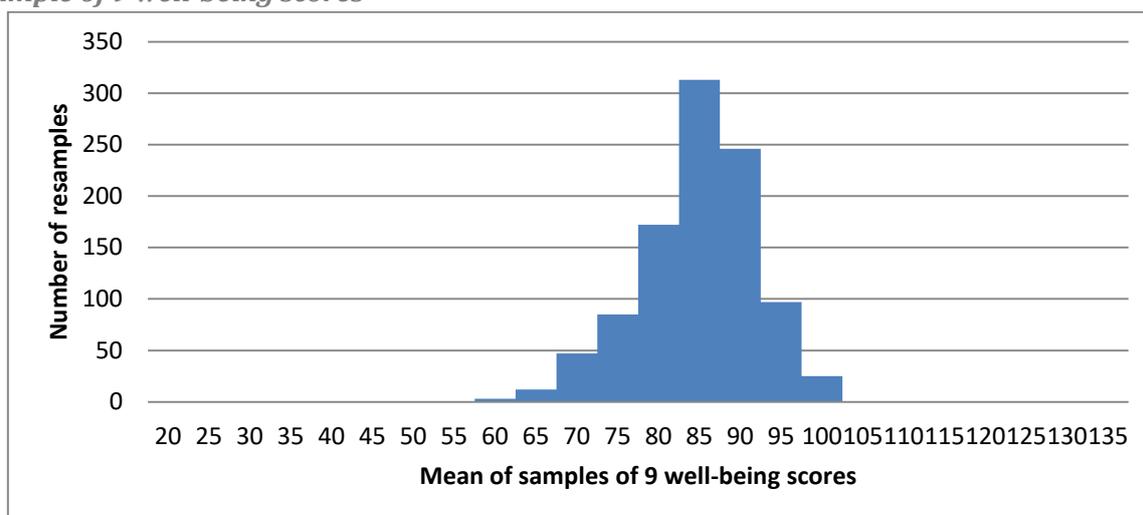

The sample of wellbeing scores on which this is based were: 96, 100, 35, 95, 97, 99, 50, 95, 98.
Sample average = 85

Despite the fact that this sample is meaningless because it is likely to be biased, let's see what happens if we imagine it was a random sample from some larger group. Using the same methods as above this leads to a confidence level of 100% for the population average being more than 50, and a 95% confidence interval for the population average of 69 to 97.

Let's see what we can infer from the 12 resamples represented by the bar centred on 65. These are 20 units below the guessed population average (based on the sample) of 85. This suggests - using the argument above - that the probability of a sample being 20 units *below* the population average is 12/1000 or 1.2%, which in turn suggests that the probability of our sample with an average of 85 being 20 units below the real population average is also 1.2%. Which suggests a 1.2% probability the population average is 20 units *above* 85 - i.e. 105. So there's a 1.2% probability that the real population average is 105. But ... as the measurement scale has a maximum of 100 this probability should really be 0%: it's impossible! This possibility is not equally likely to the other possibilities (Assumption 3): its prior probability is zero.

Now try the 47 resamples in the bar centred on 70, 15 units below the guessed population average of 85. This suggests, if we use the argument above, that there's a 4.8% chance that the average of the real population is 15 units above 85, or 100. But if the population average is 100, and no individual value can be more than 100, this means that every number must be 100 and the probability of any other average is zero! Which means that the distribution can't be a similar shape if the average were 100. Assumption 5 also fails.

Assumption 4 also fails because the distribution is not symmetrical. The average is 85, the lowest bar is 25 below this (60), but the highest is only 15 above (100).

In practice, if the distribution is symmetrical like Fig 5 (Assumption 4), it is probably reasonable to assume Assumption 5 because the symmetry suggests that population values above the sample average are likely to lead to a similar sampling pattern to those below the average, whereas in Fig.

11 the asymmetry suggests this is not the case, and 100 is certainly a special case. So Assumption 5 can (perhaps) be ignored. Symmetrical distributions are, in practice, commoner than you might suppose because there is a well known tendency, described by a famous theorem of statistics[15], for distributions of averages to tend to be symmetrical, particularly for larger samples.

*This means the four assumptions to check for the bootstrap argument are (1) the sample should be random or similar to a random sample, (2) the sample should be large enough to give a reasonable guessed population for the bootstrap simulation, (3) the prior probabilities of each possibility should be equal, and (4) the distribution should be roughly symmetrical.*

However, despite its lack of symmetry, Fig. 11 does seem plausible as a confidence distribution like Fig. 6, *because of its asymmetry*. (A symmetrical distribution would not make sense because it would include possibilities over 100.) The fact that the argument above does not fit, does not, of course, mean that the conclusion is wrong just as the invalidity of one argument for the existence of God[16] does not prove that God does not exist.

There are two other ways we could check the bootstrap argument. The first is the check it against standard methods. The bootstrapped 95% confidence interval for Figure 6 is between 49 and 72; the equivalent from the conventional method using probability theory in SPSS is 45 to 75. For Figure 8 the bootstrapped interval is 69 to 97; the equivalent from SPSS is 66 to 104. bearing in mind that we are in the business of trying quantify the accuracy of guesswork by two very different methods, I would say these results are close. Both bootstrap intervals are narrower: this may be because the small samples inevitably underestimated how variable the sample were because small sample are unlikely to include the extremes in the populations.

The other check we could do is to do some experiments to see how accurate bootstrap confidence intervals are with random samples from given populations. In general, bootstrapping does well in these comparisons, but various elaborations on the bootstrap theme - which inevitably add complexity to a beautifully simple idea - tend to do a bit better.

My suggestion is simply to check [Assumptions 1- 4](); use bootstrapping if these are a reasonable approximation, otherwise think again or accept that the answers will only be very rough.

## Relation to conventional concepts and methods

I've only used jargon where I think it's helpful, and I've ignored some standard techniques and concepts where I think there are better alternatives. I've also extended the idea of confidence intervals to derive tentative probabilities for hypotheses - which is not part of conventional statistics. However, if you are reading articles using standard concepts and jargon, or if you want to use a computer package to analyze your data, you will need to know how the concepts and methods introduced above relate to the standard ones. If you aren't doing either of these I'd suggest ignoring this section.

The main omission is the jargon around *Null hypothesis significance tests* (*NHST*s, or *significance tests* or simply *tests*), and the mathematical background to these. I have called this approach to analysis [Approach 1: Is the data consistent with a baseline hypothesis?]() My baseline hypotheses are conventionally called *null hypotheses*, and *p values* are also called *significance levels,* except that the latter term is sometimes reserved for conventionally set levels such as 0.05 or 0.01: a result is then described as *statistically significant* or simply *significant* if *p* is less than the 0.05 (or other chosen level). A significant result means that the data is unlikely to have arisen from the null (baseline) hypothesis so this is likely to be false and an alternative hypothesis must be true. And the lower the p value the stronger the evidence against the null hypothesis is. In many fields of research NHSTs are very much the norm, despite widespread criticism going back at least 50 years (Morrison & Henkel, 1970; Nickerson, 2000; Nuzzo, 2014; Wood, 2014, pp. 5-6): they are very widely misunderstood, and they don't tell you how likely the alternative hypotheses are. They have their place, but Approach 2 is often far more useful.

A significance test in the singular is simple a recipe for doing this is in a particular set of circumstances. There are many such recipes - including the t test, ANOVA, the $\chi^2$ test, and so on. Each of them uses probability theory to work out p values instead of using simulation experiments like we have above. As a very simple example of probability theory, if you toss two coins the probability of getting two heads is 25%; if you toss three coins the probability of getting two heads is 37.5%. These can be worked out by pure thought, which can be formalised in mathematical recipe or formulae, starting from the assumption that heads and tails are equally likely when a coin is tossed[17]. (The simulation approach to this - see the probability tab in the spreadsheet at http://woodm.myweb.port.ac.uk/SL/resample.xlsx - will give you an answer that's close to 37.5% but it won't be exact.) The maths behind this is called the binomial distribution and is the basis of a significance test called the sign test. The t test, ANOVA, and so on, all have formulae based on probability theory behind them, but more complicated than the binomial distribution. And there are an awful lot of these tests, each with their own complicated mathematical recipe. Using simulation approaches like the shuffle test and bootstrapping makes life an awful lot simpler!

The simulation approach gives you an answer without any complicated maths. For example, the t test could be used instead of the shuffle tests for the example above. The answer from the shuffle test is about 2%; the p value produced by the "unpaired t test", carried out by the statistical package SPSS, is 2.4%. These are slightly different: partly because the shuffle test uses random numbers and the answer will be different each time, and partly because they ask slightly different questions and make slightly different assumptions (arguably the shuffle test is more sensible than the t test[18]).

But, you may be thinking, does this mean I have to know jargon like "unpaired t test" so that I know which technique to use? Well, no, actually you don't. If you are using a computer package, like SPSS, to compare two means or anything else, the statistics package will help choose the appropriate test, do the appropriate sums and give you the p value. And if you are reading an article just ignore the details of the test used. All you need to know is what the p value means.

Even if you don't use the shuffle test, or another simulation approach, they should provide a mental image of what's going on and what p values mean. P values are probabilities, and all probabilities can be simulated by setting up an appropriate scenario and running it through lots of time.

Confidence intervals are conventionally included in introductory texts and software packages, and are widely used in some areas of research (e.g. medicine) but not in others (e.g. business). The usual method of calculation uses probability theory (the t distribution is often part of the method) instead of simulation methods, but the underlying concept is the same.

An important difference between my approach here and the usual statistical story is that there is no equivalent to Approach 2: estimating probabilities for hypotheses. The idea of confidence is not extended beyond confidence intervals to more general questions like whether the average wellbeing score of vegetarians is more than 50. This strikes me as being very odd because it is a very obvious useful thing to do. Part of the problem is that in the standard story confidence is seen as being different from probability. This seems so unreasonable to me that I can't really make sense of the reasons, so I won't attempt to explain. If you google it and get confused, you have my sympathy. But it is important to remember that confidence levels are very tentative probabilities (see above): you should always remember they may be misleading.

## How to estimate tentative probabilities for hypotheses from confidence intervals or p values

This raises the question of how conventional methods could be adapted to, for example, estimate the probability of the average wellbeing score of vegetarians being more than 50 based on the sample in Figure 4. What if we just know that the confidence interval extends from 49 to 72?

This is easy to do because confidence intervals can be (and usually are) worked out from a mathematical function which is built into Excel: the t distribution. In practice it is easier, and usually makes little difference, to use the normal distribution which is the bell shape curve which so many histograms approximate to (e.g. Figures 3, 5 and 7 above). It is quite easy to use the confidence limits to find the right shape normal distribution and read off the corresponding probability. I've set this up on the spreadsheet http://woodm.myweb.port.ac.uk/CLIP.xls from where Figure 12 below comes. According to this spreadsheet the probability of the overall average being more than 50 is 96%; according to the bootstrapping method above the probability is 95%. This is as close as we should expect remembering that the bootstrap method will give a different answer each time, and the rationale behind the two approaches are very different. Using the t distribution also gives 96% to the nearest whole number: the difference between using the t and normal distributions is only 0.29%.

*Figure 12. Using the normal distribution to estimate confidence levels*

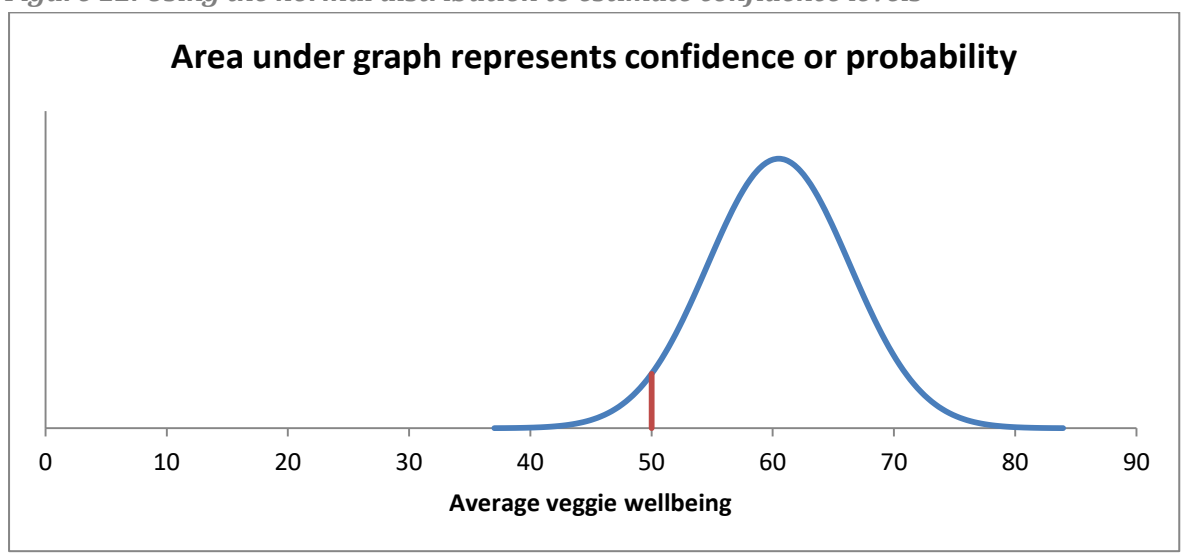

It is worth comparing Figure 12 with Figure 6. Figure 6 was produced using the bootstrap method without any elaborate mathematics. Figure 12 is based only on the two confidence interval limits from Figure 6 (49 and 72), with the rest of the curve worked out by probability theory (of the normal distribution). The two are remarkably similar.

In many fields of inquiry confidence intervals are not routinely produced. All we are given are p values. (Crazy in my opinion, and the opinion of many others.) However, making suitable assumptions, these can also be used to reverse engineer confidence levels. For example, the p value for the baseline (null) hypothesis that vegetarians and omnivores have equal levels of wellbeing based on the sample of 60 in Figure 3 was 2% from the shuffle test. What if we just know that the p value is 2% and that the difference of the sample averages is 10 in favour of the veggies?

Again, I've set this up in http://woodm.myweb.port.ac.uk/CLIP.xls which gives the answer that probability of the difference being positive is 99% and the probability of the veggie average being more than 5 more than the omnivores is 88% (using the normal or t distribution). The corresponding answers from the bootstrap method above are 99% and 87%. Very close!

*However, please remember that probabilities produced by these methods are very tentative. In particular, they rely on the assumption that the distributions involved (like Figures 3, 5 and 7) are symmetrical like the normal (and t) distributions.*

In the next section I'll look at how this all relates to some published research.

## Deciphering published research reports

The statistical analysis used in most published research articles tends to be on the complicated side and involve an array of concepts going way beyond those discussed here. Usually this is for good reasons, but I do sometimes wonder - especially in the social sciences - whether this is actually necessary or if it's just to impress reviewers and get the article published.

Typically data is used to calculate statistics like averages, proportions, correlations or more esoteric measures. The first task is to be clear about what the statistic is and what it means.

The "How sure are we?" question is then usually quantified by one or both of two methods:

- P values, typically calculated using computer packages. The null, or baseline, hypothesis is often unstated: it is important to be clear exactly what the hypothesis on which the p values are based is.
- Confidence intervals.

Occasionally Bayesian methods are used but this is rare in most fields of inquiry.

The section above deals with how these p values and confidence interval relate to the simulation methods presented above. I also suggested that it may be useful to go beyond the results presented in the article to derive confidence levels or probabilities for hypotheses other than intervals. I'll illustrate this with an example which illustrates several of the issues.

Articles in medical journals often cite both confidence intervals and p values. To take a recent example, Wallis et al (2017), in a "large, population based, matched cohort analysis found small differences in surgical outcomes between patients treated by female and male surgeons, with the former having a small but statistically significant decreased risk of short term postoperative death." In part of their analysis, cited in the abstract, they found that "patients treated by female surgeons were less likely to die within 30 days (adjusted odds ratio 0.88; [95% confidence interval] 0.79 to 0.99, P=0.04), but there was no significant difference in readmissions or complications."

This sounds important: being cut open by a women is less risky than being cut open by a man. But how much less risky and how sure are we that the research is right, that it will apply to future operations as well as those in the past on which it is based?

The first thing is to be clear about is the statistic quoted: the odds ratio. Wikipedia is a good starting point if you are in doubt, but I will give a summary here. The research paper gives the actual figures, but it is easier to see how it works with smaller numbers and exaggerated death rates. Suppose that 4 out 10 patients of female surgeons died so 6 survived, and 8 out 10 patients of male surgeons died leaving only 2 survivors. Then the odds of death for female surgeons is 4:6 or 4/6 as a fraction. The corresponding figure for male surgeons is 8:2 or 4 implying that there are four times as many dead ex-patients as living ones. The odds ratio is the ratio of these ratios which is 4/6 : 4 or 4/24=1/6 as a fraction. Which seems to imply that being operated on by a man is six times as dangerous as being done by a woman.

But ... this is not really right. The natural measure for how dangerous an operation is the risk or probability of death. For female surgeons this is 4/10 and for males 8/10, so the ratio of the risks is 40%/80% or 1/2 or 50%. In other words male surgeons are twice as dangerous, not six times. The ratio of the risks is a far more intuitive and useful measure than the odds ratio. The odds ratio tends to be used instead of the more intuitive risk ration because of its mathematical convenience (Nurminen, 1995), but it's definitely less convenient from the reader's perspective.

However, in the actual research the death rates are much smaller (about 1%), and difference between the odds ratio and the risk ratio is also smaller. To the nearest whole number the risk ratio corresponding the result cited above is the same as the odds ratio: 88%. So we can interpret this as a risk ratio: the risk of death from the knife of a female surgeon was 12% less than the risk from a male. Which sounds bad, but it is important to remember that the overall risk of death was low (1%) with the difference between the death rates of males and females being only 0.12%, and that the matching process to ensure that the comparison between males and females is fair and the adjustments for other variables (implicit in the phrase "adjusted" odds ratio) may have introduced small biases despite the best efforts of the statistical analysis (for example, only 5% of the data on male surgeons was in the matched cohort on which the results were based, which obviously may have introduced biases).

The 95% confidence interval for the odds or risk ratio (79% to 99%) gives a good feel for its accuracy from the sampling point of view. Would we get the same answer from another sample? Probably not, and the confidence interval gives a good feel for how much variation to expect. The confidence intervals were not derived by the bootstrap method described above, but a similar simulation method would have given a similar result, and the method should give you a feel for what the confidence interval is telling us. Bearing in mind that there were more than 100 000 surgeons in the matched cohort, the width of this confidence interval may surprise you. Despite such a large sample, there is still an error of about +/- 10% as defined by the 95% confidence limits. There may be a general lesson here: conclusions based on samples are often less reliable than we intuitively assume.

We are also told that p=0.04 or 4%. The unstated baseline, or null, hypothesis here is that death rates of female and male surgeons are the same and any differences found in the sample are just due to chance. The chance in question is 4%, which is less than the conventional cut-off of 5%, so is described as a "statistically significant decreased risk of short term postoperative death". On the other hand we are told that "there was no significant difference in readmissions or complications". Table 2 in the paper shows that there is actually a small difference (risk ratio 96% in each case), but the p value is 20% in one case and 10% in the other so this is deemed not significant.

There are two problems with the way this information is presented. First it's misleading because the adjective "significant" seems to refer to the difference or reduction in the risk, whereas it should refer to the evidence for the difference or reduction. Stronger evidence (e.g. with an even larger sample) would lead to a more significant result but the size of the difference could be the same. More accurate phrases would be "statistically significant *evidence for* decreased risk of short term postoperative death" and "no significant *evidence for a* difference in readmissions or complications". However, this slightly more convoluted phrasing, in which the word significant refers to the evidence, is rarely, if ever, used, which encourages the widespread misinterpretation that a significant result is a big or important difference, and that no significant difference means no difference. (A more precise, but even more convoluted, wording would be "significant *evidence against the hypothesis that there is no* difference in readmissions or complications".)

The second problem is that the p value does not directly tell us how likely it is that female surgeons are less likely to kill a patient than male ones.

Both problems can be resolved by estimating a tentative probability, or confidence level, for the risk ratio being less than one, i.e. for the female surgeons having a lower death rate than the male ones. Using the spreadsheet at http://woodm.myweb.port.ac.uk/CLIP.xlsx, starting from either the p value (0.04) and the value from the sample (0.88), or from the 95% confidence interval (0.79 - 0.99), this probability comes to 98%. For the normal distribution to be a reasonable approximation (see above) the confidence interval must be reasonably symmetrical - which it is because the middle of the interval is 0.89 which is close to the sample value of 0.88. We can, tentatively, be 98% sure that female surgeons do have a lower death rate than male surgeons, although the difference is only a small one. I would rewrite the conclusion in the abstract as

> Patients treated by female surgeons were slightly less likely to die within 30 days (tentative probability 98%; adjusted risk ratio 0.88, 95%, confidence interval 0.79 to 0.99) ...

Replacing the p value with the tentative probability avoids the misleading use of the word "significant", and gives a clear, although tentative, probability for the conclusion. And the risk ratio is more interpretable than the odds ratio.

We can do the same for results that are not significant: readmissions and complications. Taking readmission to hospital as our example, the equivalent statistics were risk ratio=0.96, 95% confidence interval: 0.91 to 1.02, P=0.20. Using the spreadsheet  gives a tentative probability that female surgeons have a lower readmission rate of 89% (using the confidence interval) or 90% (using the p value: the difference is likely to be a rounding error). This avoids the implication behind the phrase "no significant difference" that there is no difference: there is a small difference and the evidence for the direction of the difference is weaker although be no means negligible: a tentative probability of about 90% rather than 98%.

Alternatively, we may feel that, given the likely inaccuracies of the matching and adjustment processes, a difference in the readmission rates of less than 10% is not meaningful. The [spreadsheet](spreadsheet) can be used to estimate the tentative probability of the difference being less than 10%: this comes to about 98% (by estimating the probability of the risk ratio being <0.9 and >1.1 and subtracting these two probabilities from 100%). This represents strong evidence that there isn't a meaningful difference.

Very similar points apply to many other research papers, perhaps the majority in some fields. In an earlier article (Wood, 2014) I used the example of some research which found that "life expectancy was 3.9 years longer for Academy Award [Oscar] winners that for other, less recognized performers (79.7 vs. 75.8 years, *p* = 0.003)" (Redelmeier and Singh, 2001: 955). No confidence interval is given, so readers get no feeling for how accurate and reliable the result is likely to be. The p value is equivalent to a confidence, or tentative probability, level of 99.85% which gives a far clearer indication of the level of confidence we should have in the result.

However, according to a subsequent article by Sylvestre et al (2006) published in the same journal five years later

> ...the statistical method used to derive this statistically significant difference gave winners an unfair advantage because it credited an Oscar winner's years of life before winning toward survival subsequent to winning. When the authors of the current article reanalyzed the data using methods that avoided this 'immortal time' bias, the survival advantage was closer to 1 year and was not statistically significant.

Which, of course, implies that the new tentative probability for the hypothesis that Oscar winners live longer would be lower.

Without going into the details of the criticism of the original article, the fact that it took five years for the journal to admit that the original headline grabbing conclusions were problematic should alert us to the subtlety of the concepts used in many statistical investigations, and the importance of recognising the possibility of flaws in apparently secure research.

The articles we've looked at so far in this section have been from medical journals. The style of statistical analysis in some other disciplines makes deciphering their conclusions harder. In another article (Wood, 2013) I looked at an article in a management journal, chosen because it was relatively simple and clearly written (compared to other articles in the journal). The article (Glebbeek and Bax, 2004) "tested the hypothesis that employee turnover and firm performance have an inverted U-shaped relationship: overly high or low turnover is harmful" (p. 277), with the optimum level of turnover lying somewhere in the middle. Despite this hypothesis the article does *not* include a graph to show the reader the inverted U shape in question, and the credibility of the hypothesis is checked by citing *two* p values, neither of which is significant, all of which makes it extremely difficult to understand or check the authors' conclusions. Fortunately the authors made their data available to me which meant that I could draw the graph from the data - which shows a rather scattered inverted U-shape - and use a bootstrapping method to derive a tentative probability for the inverted U-shape hypothesis of 65% (Wood, 2012).

Glebbeek and Bax (2004) also tested a straight line model of the relationship between employee turnover and performance. This is a very common type of analysis so is worth examining detail. The relevant table (Table 2) gives a "standardized" regression coefficient of -0.27** where the ** indicate that p<0.01. The standardization makes the regression coefficients very difficult to interpret, so they explain in the text that this coefficient "indicates that a 1% increase in turnover equals a loss of 1780 Dutch guilders [this was before the introduction of the Euro] .... From a management point of view, this is quite substantial" (p. 283). The figure 1780 is the unstandardized regression coefficient: it would seem to me to be sensible to tabulate this number, not the standardized coefficient. As is standard in management research, confidence intervals are not given, although they are the obvious way of quantifying the uncertainty in the result: the 95% confidence interval extends from -3060 guilders (per 1% increase in staff turnover) to -500 (Table 2 in Wood, 2013). This indicates a considerable degree of uncertainty. We can also use http://woodm.myweb.port.ac.uk/CLIP.xls to calculate a tentative probability for this figure being negative (for higher staff turnover to lead to lower performance): this comes to 99.7%.

This example is typical of much research in the social sciences in that the statistical results presented are almost designed to be obscure. Glebbeek and Bax (2004) tested a hypothesis about an inverted U shaped graph, but no graph is shown. As evidence for the reliability of their conclusion they cite two non-significant p values which are difficult to interpret, instead of the 65% confidence level from bootstrapping. And for the linear (straight line) model they give an uninterpretable standardized regression coefficient instead of the unstandardized version (a loss of 1780 guilders for each 1% increase in turnover), and fail to quantify the uncertainty in this figure by means of a 95% confidence interval (from a loss of 500 to a loss of 1780 guilders).

## Conclusions

Statistical inferences which go beyond the sample of data studied can never be completely certain. This paper looks at two broad approaches to assessing the degree of confidence we should have in such inferences. The first approach is usually called null hypothesis significance testing, although I prefer to call it checking if the data is consistent with a baseline hypothesis because this seems a clearer description. The second approach aims to estimate a "tentative probability" for hypotheses or conclusions of interest. In the illustrative example used above, the first approach yields p value of 2%, indicating that the data is not very likely to have occurred if the baseline hypothesis, of equal average levels of wellbeing in the populations of vegetarians and omnivores, were true. The second approach yields a tentative probability of 99% for the population average wellbeing for vegetarians being higher than that for omnivores.

The second approach seems preferable because it produces a clear probability for the conclusion of interest, whereas the first approach provides no such probability. On the other hand, the probability produced is necessarily tentative. At present, the second approach is never used, except in a Bayesian framework. I think that in many contexts it makes much more sense than the first approach. Qualifying the conclusion that "patients treated by female surgeons were less likely to die within 30 days" (Wallis et al, 2017) by stating a tentative probability for the conclusion of 98%, is surely clearer and more informative than "P=0.04"?

In the presentation of the two approaches I have tried to keep the conceptual framework as simple as possible in several ways. First, there are only two approaches. Second I've tried to simplify the jargon as much as possible, and reduce the possibility of misunderstanding. For example, I've avoided the word "significant" (often taken to indicate a large or important effect), I have treated probability and confidence as synonymous because the distinction seems an unnecessary distraction, and instead of presenting a number of different mathematical models for different circumstances, I have used simulation methods of far greater generality, all implemented by the same spreadsheet which is available on the web. Simulation methods have the further advantage that understanding their rationale does not require elaborate mathematics. They are, in this sense, far simpler than their mathematical equivalents.

Wood, M. (2014). P values, confidence intervals, or confidence levels for hypotheses? https://arxiv.org/abs/0912.3878.

# Appendix: What are probabilities and where do they come from?

The definition of probability at https://en.oxforddictionaries.com/ is "the quality or state of being probable; the extent to which something is likely to happen or be the case." This is made more precise in a mathematical context: "the extent to which an event is likely to occur, measured by the ratio of the favourable cases to the whole number of cases possible."

These two definitions neatly summarise the dilemma for probability theorists. The general definition covers events which may or may not happen, *and* assertions that may or may not be true, whereas the mathematical definition covers *only* events which may or may not happen.

The difficulty with the general definition is that the assertions are assumed to be either true or not, so the probability can only refer to our beliefs about their truth which is, according to conventional statisticians, a horribly subjective idea not amenable to rigorous analysis. So probability must be restricted to events and can't be applied to hypotheses which may or may not be true. As the editor of a journal to which I submitted a paper on a similar topic wrote "I am going to flip a coin, so the probability that it will come up Heads is .5. I flip the coin, look at it, but do not show it to you. To you, it still feels like the probability of Heads is .5, even though I know the answer. But just because it feels good does not mean there is a good reason for thinking of the probability as .5."

But he (he was male) is failing to see the point of mathematics, which is to invent concepts which are useful and then check they are consistent and work out their properties - all of which contribute to their usefulness. So if it feels good, and it's obviously a useful concept, then it deserves to exist. Mathematicians tend to be a bit anal about things like this: they want, firstly, to show how everything follows from indisputable axioms so that their conclusions follow with complete certainty from their assumptions, and secondly, they want a clear way of interpreting probability in the real world.

It's the second of these that's the problem. The approach to statistics called Bayesian statistics (the rudiments of which are explained above) works with the idea of probability as a degree of belief. However, as the degree of belief is usually interpreted in a subjective sense, this school of statistics is usually treated with suspicion.

*My suggestion here is to sidestep all these issues by defining the probability of any characteristic in a particular context by imagining a set of equally likely possible worlds, some of which possess the characteristic and some of which don't. The probability is then simply the proportion of possible worlds which exhibit the characteristic.*

This taps into the intuitive idea of probability derived from games of chance. For example, if you deal a card from a well-shuffled pack of 52 cards, what's the probability of the card dealt being a spade? Because 13 of the 52 cards, or 25%, are spades, the probability in question is obviously 25%. The rationale behind this is very simple. There are 52, equally likely, possibilities for the card to be dealt, 13 of them are spades, so the probability of a spade is 13/52 or 25%.

The equally likely bit is important. If I enter for next week's lottery, there are two possibilities: either I win the jackpot or I don't. But they aren't equally likely, so we can't say the probability of my winning the jackpot is 1/2 or 50%. If we want to work out the probability of winning the jackpot, I need to work out how many equally likely outcomes are possibly in the lottery, and how many will lead to me winning, and then the probability is simply the second number divided by the first.

What about the probability of rain falling sometime in London tomorrow? Here there is no list of equally likely possibilities, but I've just googled the question and found an estimate of a 10% probability of rain from weather.com. For the day after the probability given is 60%. How can we fit this into the equally likely possibilities view of probability. We obviously need to invent 10 possible tomorrows of which just one has rain. Then the God, or whoever, throws the dice (ten sided in this case) to determine which of the ten tomorrows becomes the real tomorrow.   The possible tomorrows are, of course, an invention: they are a mental model to make sense of the 10% probability in the same terms as a game of chance. If the probability were 15%, we would obviously need to imagine 20 tomorrows, of which three have rain, or 100 of which 15 have rain.

But why bother? Normally, we wouldn't if we can handle the meaning of these probabilities without any trouble. But there are two reasons for indulging in these mental gymnastics.

First, it gives us a way of interpreting a probability. What does the 10% chance of rain tomorrow actually mean? Imagining 10 possible tomorrows gives us a way of linking the 10% to intuitions about games of chance.

Second it gives us way of dealing with more complicated situations. What about the probability of rain both tomorrow and the day after. For each of the 10 possible tomorrows we could imagine 10 possible days after tomorrows, on six of which it rains (reflecting the 60% probability): this gives us 100 possibilities for the two days. How many of these possibilities involve rain on both days? The answer is obviously 6/100 or 6% (Figure 13).

*Figure 13. 100 possible tomorrows and the day after assuming statistical independence*

| **RR** | FR | FR | FR | FR | FR | FR | FR | FR | FR |
|--------|----|----|----|----|----|----|----|----|----|
| **RR** | FR | FR | FR | FR | FR | FR | FR | FR | FR |
| **RR** | FR | FR | FR | FR | FR | FR | FR | FR | FR |
| **RR** | FR | FR | FR | FR | FR | FR | FR | FR | FR |
| **RR** | FR | FR | FR | FR | FR | FR | FR | FR | FR |
| **RR** | FR | FR | FR | FR | FR | FR | FR | FR | FR |
| RF | FF | FF | FF | FF | FF | FF | FF | FF | FF |
| RF | FF | FF | FF | FF | FF | FF | FF | FF | FF |
| RF | FF | FF | FF | FF | FF | FF | FF | FF | FF |
| RF | FF | FF | FF | FF | FF | FF | FF | FF | FF |

Weather conditions on first and second day: R = rain; F = Fine (no rain). RR in bold. Statistical independence is explained below.

But... this is unlikely to be right because rain the day after tomorrow is probably more likely if it rains tomorrow. In England, when it starts raining it tends to keep on raining! Let's suppose weather.com gave us a probability of rain the day after tomorrow of 90% if it rains tomorrow, and 50% if it doesn't. (I made these figures up; weather.com does not give this much detail.) Figure 2 shows that 9 of the 100 possibilities involve rain on both days, giving a probability of 9%.

*Figure 14. 100 possible tomorrows and the day after, taking account of dependence of weather on second day on that of first day*

| **RR** | FR | FR | FR | FR | FR | FR | FR | FR | FR |
|---|---|---|---|---|---|---|---|---|---|
| **RR** | FR | FR | FR | FR | FR | FR | FR | FR | FR |
| **RR** | FR | FR | FR | FR | FR | FR | FR | FR | FR |
| **RR** | FR | FR | FR | FR | FR | FR | FR | FR | FR |
| **RR** | FR | FR | FR | FR | FR | FR | FR | FR | FR |
| **RR** | FF | FF | FF | FF | FF | FF | FF | FF | FF |
| **RR** | FF | FF | FF | FF | FF | FF | FF | FF | FF |
| **RR** | FF | FF | FF | FF | FF | FF | FF | FF | FF |
| **RR** | FF | FF | FF | FF | FF | FF | FF | FF | FF |
| RF | FF | FF | FF | FF | FF | FF | FF | FF | FF |

Weather conditions on first and second day: R = rain; F = Fine (no rain). RR in bold.

The difference between these two is that in Figure 13 we assume *statistical independence*: this means that what happens on the first day does not influence our assessment of the probabilities for the second day. This is not true for the Figure 14 where our assessment of what is likely to happen on the second day is *dependent* on what has happened on the first day.

This may all seem a bit fanciful, but the idea of probability is to see how likely various possibilities are, so the idea of making all the possibilities explicit is almost a logical prerequisite for this process. I used the phrase "possible tomorrow" in the weather example, but the metaphor of a possible world is useful because it emphasises the fact that there are many factors interacting to create the weather tomorrow. Conventional statistics would talk of events which can be counted up, but tomorrow's weather is unique and there is no series of events which can be counted. A list of hypothetical possibilities is a firmer basis for an image of probability than a sequence of events.

The *equally likely possible worlds model of probability* gives us a way of estimating probabilities from first principles - as illustrated in Figures 13 and 14, and in the [explanation of Bayes theorem above](). Conventionally probabilities are viewed as numbers between 0 and 1 which are manipulated according to the laws of probability. Using the model of equally likely possible worlds avoids all this. Obviously, for complicated arguments, conventional methods are likely to be simpler because the notation is more compact, and if, for example, we were considering more than two days in our weather example, the number of worlds would quickly become unmanageable. Alternatively, Monte Carlo simulation ([see below]()) can help with complicated scenarios. But provided we don't make it too complicated, stories about equally likely possibilities provide an intuitive model to make sense of probability theory.

There is still an important distinction between games of chance like cards, and weather forecasting. In the first case, the equally likely possibilities model is *definite*. No sensible person could doubt that it: if the pack and dealer are "fair", the probability of an ace is 4/52. (The argument is, of course circular, like most definite arguments!). On the other hand the information on which Figure 14 is based is certainly not definite: it must be viewed as tentative. Probabilities vary in their degree of tentativeness (see also [above]()).

## Monte Carlo simulation

As an alternative to working out a complete list of equally likely possibilities it may be possible to use a computer to generate them at random, and then use this random collection to estimate probabilities. The results won't be exact, but they are often close enough to be useful. This is known as Monte Carlo simulation because it involves treating the world like a roulette wheel in Monte Carlo.

For example, we could get the computer to choose at random between boy and girl eight times to simulate a family of eight children. The spreadsheet at http://woodm.myweb.port.ac.uk/SL/resample.xlsx will do this: click on the Probability tab at the bottom. The spreadsheet works with 1000 simulated families: when I used it 257 or 25.7% of these had four girls. This is a lot easier than writing down the 256 (equally likely) possible sequences of children (GGGGGGGG, GGGGGGGB, GGGGGGBG, GGGGGGBB, etc) and counting up the number with exactly four girls (70 or 27.3%), and is close enough to be useful.

Professional weather forecasters use a similar method, but based on models of the atmosphere. What they typically do (http://www.metoffice.gov.uk/news/in-depth/science-behind-probability-of-precipitation) is run computer simulations of the atmosphere to see if rain occurs in the simulation. The difficulty is that some of the details will be uncertain, so they run lots of different simulations with different detailed starting points. Each simulation represents a possibly tomorrow, and if each one can be regarded as equally likely, the probability of rain could be worked out as the proportion of possible tomorrows which are wet. In practice, it may not be quite this simple, but this is the basic idea.

## Notes

[1] There is a discussion of probability in the Appendix.

[2] I have done a similar experiment in many lectures on statistics, but I don't leave the result to chance or telepathy: I cheat by telling the "receiver" which card the transmitter will be looking at.

[3] This is roughly the result of votes in lectures where I have done a similar "experiment". At the time I thought telepathy was impossible, so, if cheating is ruled out, the only viable explanation was guessing however many cards were in the pack. Since then I have changed my mind. When I was doing these experiments I was just thinking of the traditional five senses - sound, sight, touch, smell, and taste. But there are obviously other senses - what about heat and the magnetic sense that some birds are believed to use to navigate, for example? And now (2017) the idea of people having implants which are the equivalent of mobile phones allowing them to communicate with others with similar implants seems entirely feasible. How do I know that Annette and Peter have not got such enhancements fitted? All of which makes telepathy seem less outlandish and more plausible. And, coming to think of it, the advantages of the surreptitious communication that such implants would enable are such that it seems entirely plausible that evolution by natural selection might have come up with something similar. So ... now I think I would change my mind with more than 1000 cards. With a one in 50 chance of guessing correctly, I think I would favour guessing as the more likely explanation. But with a one in a 1000 000 chance I would definitely favour telepathy (i.e. communication by some sense unknown to me).

[4] I am using the word average instead of mean because this is common usage in everyday language, and also because the Excel function for the mean is called AVERAGE.

[5] It is tempting to assume that the probability of Annette guessing is 2% so the probability of telepathy (the only alternative hypothesis assuming cheating is ruled out) is 100%-2% or 98%. However, this is completely wrong, although it is a very widespread misconception. The fallacy is that the 2% refers to the probability of Annette getting the right answer *assuming she's guessing*, whereas the probability we want is the probability of Annette guessing *assuming she's got the answer right,* and these are two completely different probabilities. In a similar way the probability of dealing a Queen *assuming the card is a picture card* is 33% (because a third of the picture cards are queens), but the probability of dealing a picture card *assuming the card is a Queen* is 100% (because all Queens are picture cards). These probabilities are completely different!

[6] This area of statistics is often (helpfully) called "statistical inference", but the inferences are usually assumed to refer to a "population". This is helpful because it gives a concrete metaphor, but unhelpful because there may not be a real recognisable population as in the telepathy story, or the veggie-omnie comparison if we include future potential vegetarians. It is more accurate to refer to inferences which go beyond the data to some, unspecified, wider context.

[7] Each member of the population is given a number and then the spreadsheet function =RANDBETEEN can be used to choose a random sample. For example if there were 567 people in a population, =RANDBETWEEN(1,567) will select such a sample. When I copied this function into 10 cells to choose a random sample of ten the numbers chosen were 242, 318, 10, 495 ...

[8] Probabilities are normally assumed to apply to *events* which may or may not happen, whereas confidence intervals refer to a measurement which is fixed but unknown. However, in ordinary discourse the word probability is used in both contexts, so the conventional statistical use seems to me unnecessarily restrictive. See also the Appendix.

[9] See the note above.

[10] I'm using the words model and story interchangeably to refer to a helpful fiction for understanding a problem. Model is the term usually employed, but here I prefer the word story because it better conveys the idea of the progression from Figure 8 to Figure 9.

[11] There are three ways in which the mathematical formula version is better than the deleted worlds story. First the formula is more compact, so once you've understood how it works, you can build it into other formula in a way that would be difficult with the deleted worlds story. Second, the number of worlds may get too large to visualise easily. We chose 200 worlds because this number can be divided exactly by both 50 and 4. If, say the prior probability was 63%, we would have needed many more worlds and the arithmetic involved would probably have deflected our attention from the underlying probability concepts. And, thirdly, if, say the prior probability was $1/\sqrt{2}$, no number of worlds would suffice because the square root of two cannot be expressed exactly as a fraction - it is irrational to use the mathematical jargon. This is not really a problem in practice because can get as close as we want, but we will never get it exactly right.

[12] As a simple example to illustrate the problem consider the probability of dealing a Queen from a pack comprising just the 12 picture cards (4/12 or 33.3%), and the probability of dealing a picture card from a pack comprising just the four Queens (4/4 or 100%). In conventional terminology the first is described as Prob(Queen given Picture) and the second as Prob(Picture given Queen). The difference is that the roles of Queen and Picture are reversed, resulting in very different probabilities. In the more succinct symbolic version these two probabilities are $P(Q \mid C)$ and $P(C \mid Q)$. (Conventional notation uses P for probability, and single letters for other variables, so, as P is used for Probability it can't also be used for Picture, so I have used C for Court instead.)

[13] Start with 50 telepathy worlds and 3 guessing worlds (from Fig. 9) to represent the prior probabilities. To keep to whole worlds, guessing need 50 copies, so we just create 50 copies so we have 2500 telepathy and 150 guessing worlds, so the probability of telepathy is 2500/2503 or 0.998801.

[14] I've described the possibilities above as possible worlds, but the possibilities might be different patients or events. Possibilities is a convenient general term, but when we are concerned with the probability of a hypothesis, we need to envisage possible worlds.

[15] The central limit theorem.

[16] One such argument starts by defining God as the most perfect being. He would obviously be better if he existed than if he did not exist. Therefore He must exist.

[17] With three coins there are eight equally likely results (HHH, HHT, HTH, HTT, THH, THT, TTH, TTT) of which three have just two heads. With more coins, or if heads and tails are not equally likely, this method becomes impracticable, so you need to think it through mathematically - the result is called the binomial distribution.

[18] The t test makes assumptions about the shape of the underlying distributions (for small samples they need to be roughly normal); the shuffle test avoids assumptions like these. There are similar problem with other tests based on mathematical probability theory: because they rely on assumptions that may not be true, the simulation based approaches are usually regarded as preferable from this point of view.